\documentclass[prl,twocolumn,superscriptaddress,groupedaddress]{revtex4}

\usepackage{amsmath}
\usepackage{amssymb}
\usepackage{amsfonts}
\usepackage{graphicx}

\begin{document}

\title{ Expansion dynamics of the Fulde-Ferrell-Larkin-Ovchinnikov
  (FFLO) state}

\begin{abstract}
  We consider a two-component Fermi gas in presence of spin imbalance,
  modelling the system in terms of a one-dimensional (1D) attractive
  Hubbard Hamiltonian initially in presence of a confining trap
  potential. With the aid of the time-evolving block decimation (TEBD) method, we investigate the
  dynamics of the initial state when the trap is switched off. We show
  that the dynamics of a gas initially in the FFLO state is decomposed
  into the independent expansion of two fluids, namely the pairs and
  the unpaired particles. In particular, the expansion velocity of the
  unpaired cloud is shown to be directly related to the FFLO
  momentum. This provides an unambiguous signature of the FFLO state
  in a remarkably simple way.
\end{abstract}

\author{J.\ Kajala}
\affiliation{Department of Applied Physics,  
Aalto University School of Science, P.O.Box 15100, FI-00076 Aalto, FINLAND} 

\author{F.\ Massel}
\affiliation{Low Temperature Laboratory, Aalto University School of Science, 
P.O. Box 15100, FI-00076 Aalto, FINLAND} 

\author{P.\ T\"orm\"a}
\affiliation{Department of Applied Physics,  
Aalto University School of Science, P.O.Box 15100, FI-00076 Aalto, FINLAND}

\email{paivi.torma@aalto.fi}
\maketitle 

Ultracold gases have provided experimental verification for several
fundamental concepts of quantum physics. The direct access to momentum
distribution and correlations via time-of-flight expansion has played
a key role in many landmark experiments.  Mapping of momentum to
position after time-of-flight revealed Bose-Einstein condensation
\cite{Anderson:1995dw}. Coherence manifesting after expansion gave
evidence of the phase of the condensate \cite{Andrews:1997uf}, and of
the superfluid - Mott insulator transition
\cite{Greiner:2002do}. Inversion of the aspect ratio of an expanding
Fermi gas revealed hydrodynamic behaviour
\cite{OHara:2002br,Menotti:2002ix}. A major goal is to observe the
elusive Fulde-Ferrel-Larkin-Ovchinnikov (FFLO) state which is at the
heart of understanding the interplay between superconductivity and
magnetism. While experiments in solid state systems
\cite{Radovan:2003gl,Bianchi:2003jm,Casalbuoni:2004fn} and ultracold
gases \cite{Liao:2010bu} are consistent with the state, an unambiguous
observation is lacking.

\begin{figure}
\includegraphics[width=0.45\textwidth]{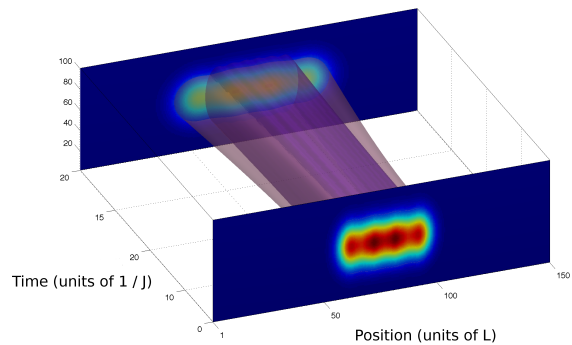} 
\caption {(Color online) Expansion of a spin-density imbalanced Fermi gas in the 1D
  FFLO state after the confining potential has been switched off.  Our
  exact simulations show an effective two-fluid behaviour: pairs and
  unpaired particles expand with different velocities. Remarkably, the
  expansion velocity of the unpaired particles is directly related to
  the FFLO momentum and provides a straightforward way for observing
  the FFLO state.}
\label{fig:schematic} 
\end{figure}

The FFLO phase is characterized by the formation of pairs with nonzero
overall momentum \cite{Fulde:1964dq,Larkin:1964uw}. This property
manifests itself through the appearance of an oscillating
(superconducting) order parameter $\Psi(r) \propto \Delta_q \exp
\left[i q r\right]$, where $q$ is the so-called FFLO momentum
($\hbar=1$) \cite{Takada:1969fo}. It is given by the mismatch between
the Fermi momenta of $N_\uparrow$ spin up and $N_\downarrow$ spin down
particles: $q=k_{F\uparrow}-k_{F\downarrow}$. In contrast,
zero-momentum ($q=0$) pairs give the conventional BCS superconductor
physics. Despite the lack of genuine long range order, the FFLO state
has been theoretically predicted to be especially stable in
one-dimensional systems
\cite{Feiguin:2007if,Batrouni:2008fwa,Rizzi:2008jk}, as well as in
quasi-1D \cite{Yang:2001iz,Orso:2007uu,Parish:2007je}. Inspired by the
first experiments in imbalanced atomic Fermi gases
\cite{Zwierlein:2006gb,Partridge:2006hx}, various methods for
detecting the FFLO state in ultracold gases have been proposed (see
\cite{Bakhtiari:2008iqa} and references therein). In this
letter we show via exact simulations that time-of-flight expansion
provides a smoking gun signature of the FFLO state in one dimension,
see Figure \ref{fig:schematic}. Expansion and consequent imaging of
densities is a widely used basic technique in experiments with
ultracold gases.

The characteristic parameters can be chosen so that the lattice model
employed is a good approximation for a continuum model of a
spin-imbalanced gas in 1D in the strong interaction limit
\cite{Yang:2001iz,Zhao:2010hk,Essler:2005uw}. 
With appropriate mapping of the parameters, in the strong interaction limit, 
our predictions are thus relevant for experiments in 1D potentials as in
\cite{Liao:2010bu}.

We consider the 1D Fermi-Hubbard Hamiltonian in presence of an overall
confining potential
\begin{equation}
\label{HH}
H = U \sum_i \hat{n}_{i\uparrow}\hat{n}_{i\downarrow} 
- J \sum_{i\,\sigma=\uparrow,\downarrow} c^\dagger_{i\,\sigma}c_{i+1\,\sigma} + h.c. 
+ \sum_{i, \sigma = \uparrow, \downarrow} V_i \hat{n}_{i\sigma} \\ ,
\end{equation}
where $c^\dagger_{i \,\sigma}$ ($c_{i\,\sigma}$) creates (annihilates)
a spin $\sigma$ particle at the lattice site $i \in \left\{1,L \right\} $, 
$J$ is the hopping constant, $U$ is the interaction strength between the two
species, and $V_i$ the strength of the trapping potential.  
For a harmonic potential $V_i=V_{ho} (i-C)^2$, where $C$ denotes the trap center, and for a box potential $V_i$ 
is specified below.

To obtain the ground state and time evolution of the system,
we have employed the (essentially) exact time-evolving block
decimation (TEBD) algorithm \cite{Daley:2004hk} with Schmidt
number $\Gamma = 150$ and simulation timestep $\Delta t = 0.02 \frac{1}{J}$.
As results from TEBD we obtain the single particle densities 
$n_{i \uparrow}(t) = \langle \Phi (t) |c^{\dagger}_{i \uparrow} c_{i \uparrow}| \Phi (t) \rangle$,
$n_{i \downarrow}(t) = \langle \Phi (t) |c^{\dagger}_{i \downarrow} c_{i \downarrow}| \Phi (t) \rangle$,
the doublon density 
$n_{\uparrow \downarrow}(t) = \langle \Phi (t) |c^{\dagger}_{i \uparrow} c^{\dagger}_{i \downarrow} c_{i \uparrow} c_{i \downarrow}| \Phi (t) \rangle$,
and the ground state pair correlation $C_{ij}$ and its momentum transform $n_k$, which are given by
\begin{equation}
\label{C}
C_{ij} = \langle \Phi | c^{\dagger}_{i \uparrow} c^{\dagger}_{i \downarrow} c_{j \downarrow} c_{j \uparrow} | \Phi \rangle,
\end{equation}
\begin{equation}
\label{nk}
n_{k} = \frac{1}{2L} \sum_{i, j}^L e^{\imath (i-j)k} C_{ij},
\end{equation}
where $\langle \Phi| | \Phi \rangle$ describes the quantum mechanical average over the state 
$\Phi$, $i$ and $j$ are lattice site indices, $\imath$ is the imaginary unit and $L$ is the size of the lattice.

We have experimented within TEBD paramater ranges
$N_{\uparrow} = 0-40$, $N_{\downarrow} = 0-40$, 
$P = \frac{N_{\uparrow} - N_{\downarrow}}{N_{\uparrow} + N_{\downarrow}} = 0.024 - 1$,
$L = 80 - 320$, $V_{ho} = 0.02 J - 0.0001 J$, and $\Gamma = 80 - 200$, where
$P$ is the polarization and $V_{ho}$ is the harmonic trapping strength. 
Many simulations were performed also using a box potential.
Single runs with the larger parameters have been done in order to 
check that the qualitative description of the dynamics stays the same when 
changing the parameters.
Indeed, the important characteristics of the dynamics are
the same in all of the scenarios. 

\begin{figure}
\includegraphics[width=0.45\textwidth]{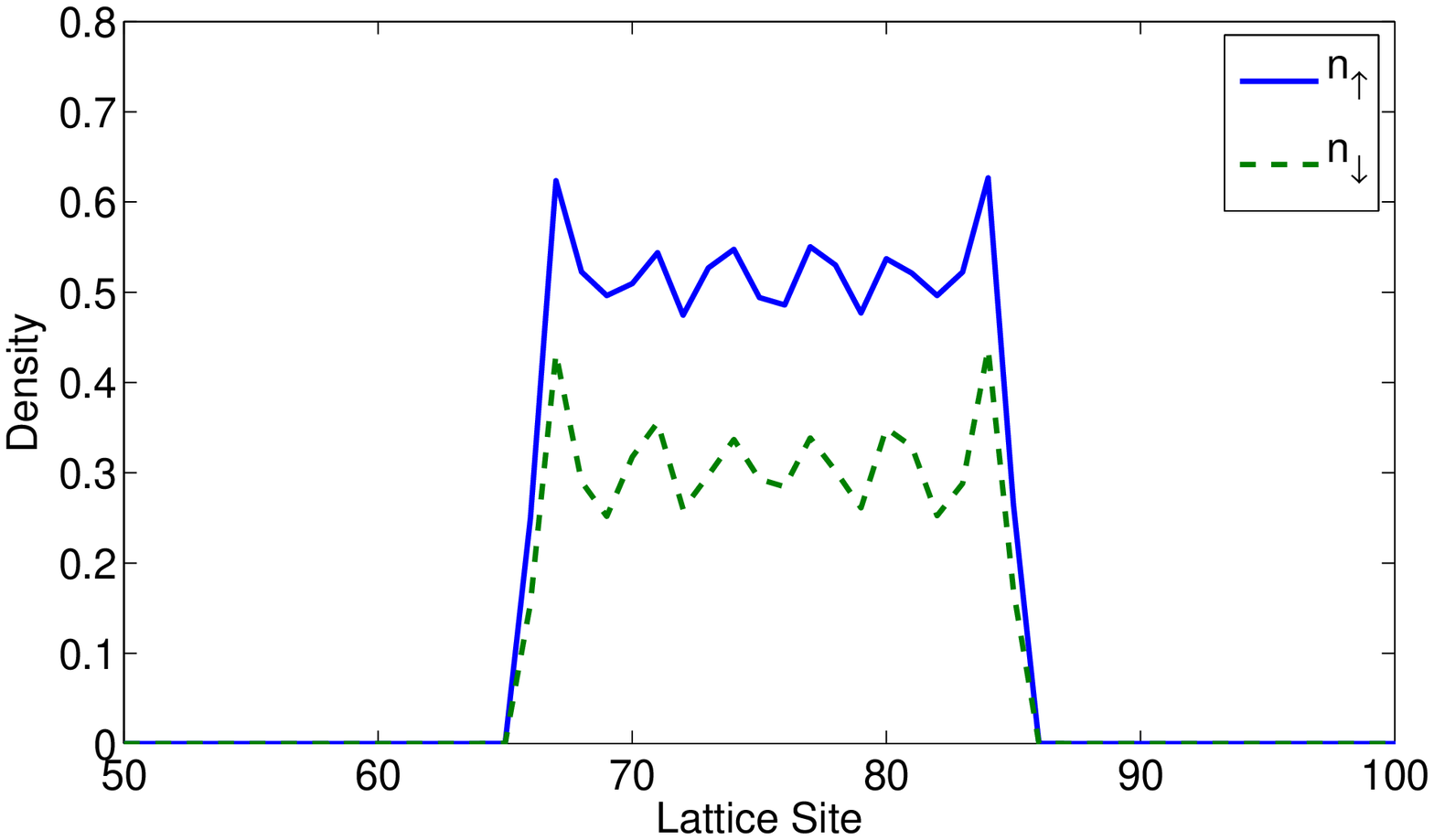} 
\includegraphics[width=0.45\textwidth]{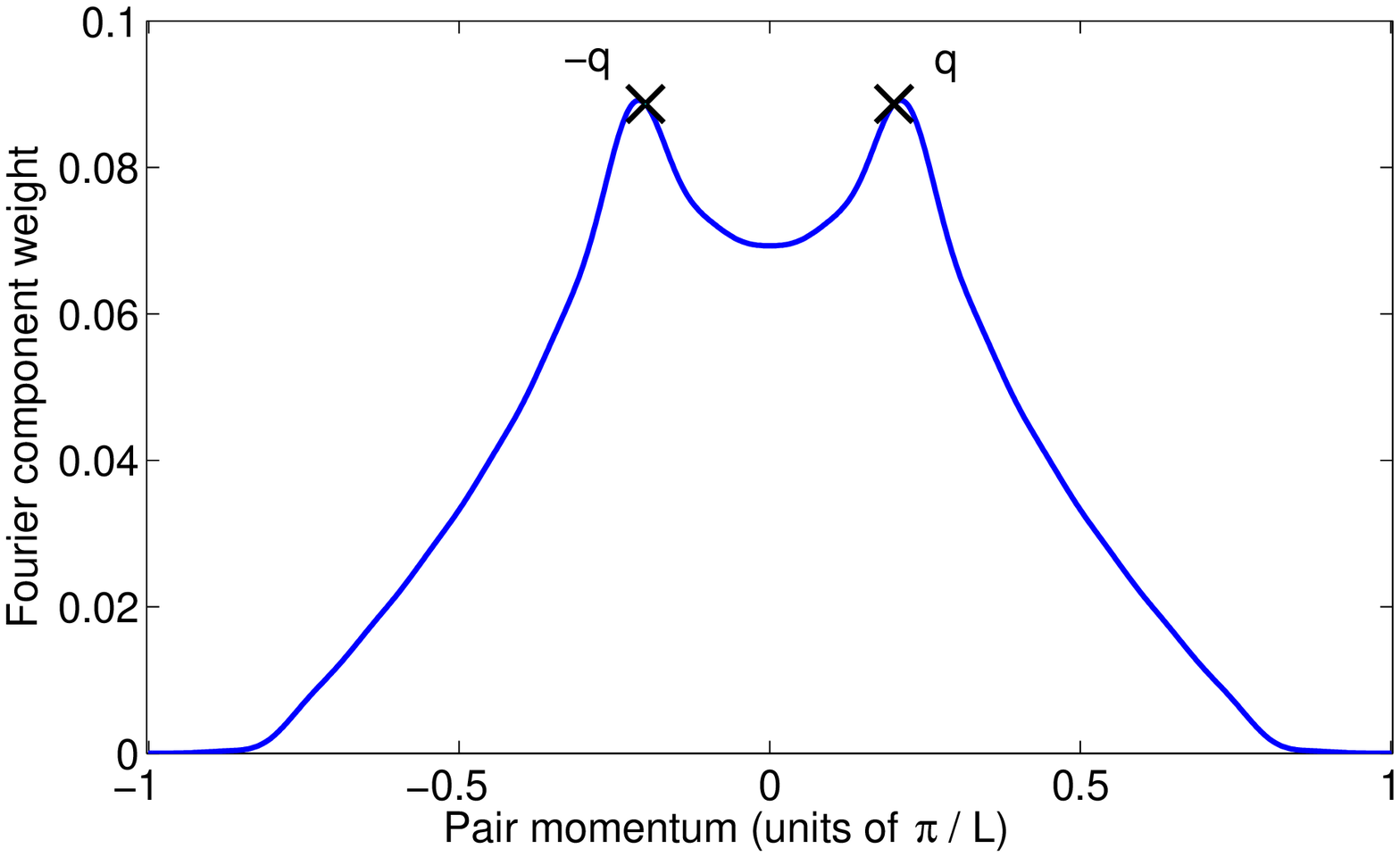} 
\caption {(Color online) \textbf{a}: The density profiles of up ($n_{\uparrow}$) and
  down spins ($n_{\downarrow}$) in the ground state when $N_{\uparrow}
  = 10$, $N_{\downarrow} = 6$, $U = - 10 J$, and there is to a good
  approximation an infinitely strong repulsive potential everywhere
  except the at the lattice sites 66-85. \textbf{b}: The pair momentum
  correlation function $n_k$ for the same state. There are peaks at
  the FFLO momenta $q=\pm (k_{F\uparrow}-k_{F\downarrow})=0.2 \pi/L$.}
\label{fig:gs} 
\end{figure}

For simplicity we first consider here
the box trap.  In Figure \ref{fig:gs} we show results for 10
up-particles and 6 down-particles, with $U=-10J$. The ground state is
a 1D FFLO state, characterized by a peak in the pair momentum
distribution $n_k$ that coincides
with the definition of the FFLO momentum $q = k_{F \uparrow} -
k_{F\downarrow}$.  Figure \ref{fig:gs}$_{\textbf{a}}$ shows the ground
state density profile in which the small oscillations characterize the
FFLO state. However, such delicate features are hard to resolve in
ultracold gas experiments, and therefore other signatures are
needed. Figure \ref{fig:gs}$_{\textbf{b}}$ displays the pair momentum
correlation function.

\begin{figure}
\includegraphics[width=0.45\textwidth]{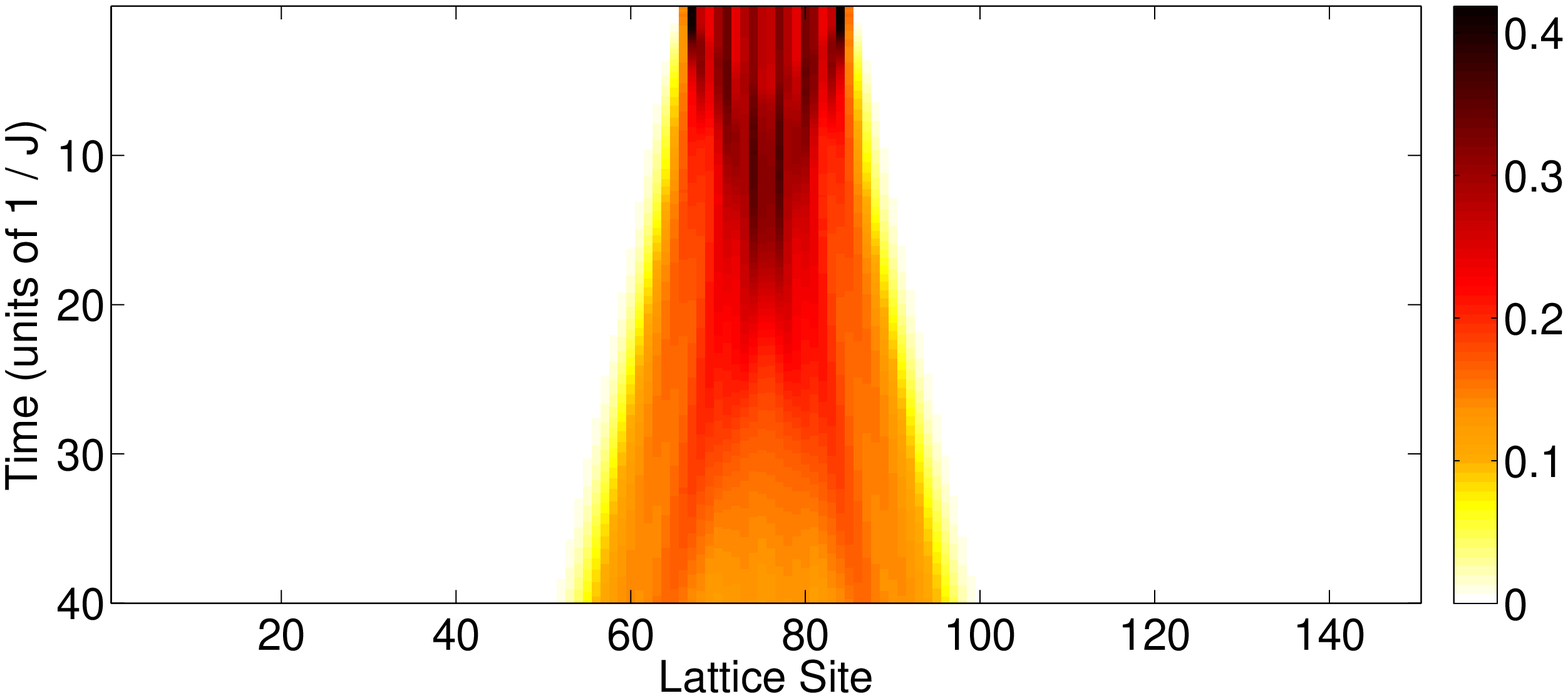} 
\includegraphics[width=0.45\textwidth]{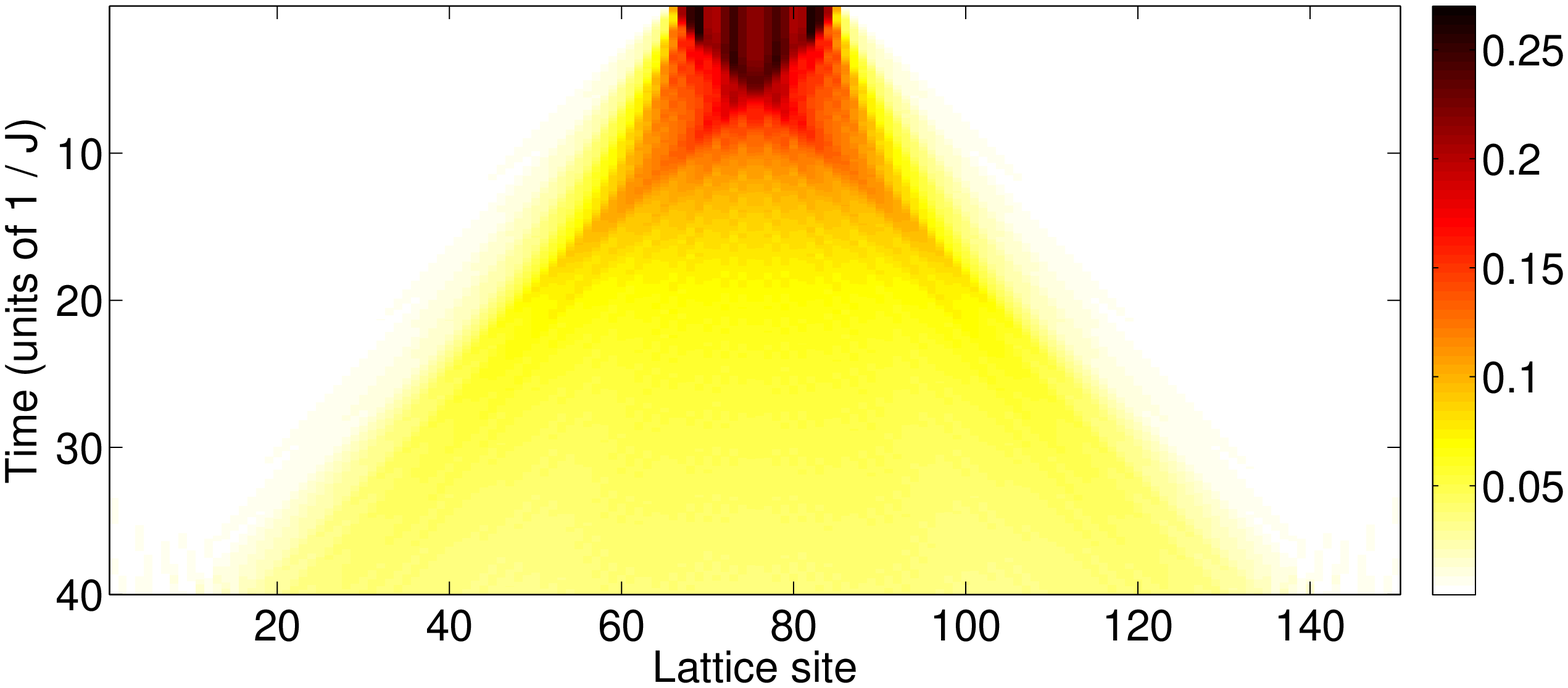} 
\caption {(Color online) \textbf{a}: The time development of the doublon density 
$n_{\uparrow \downarrow}$, corresponding to the ground state shown in Figure
\ref{fig:gs}. \textbf{b}: The time development of the unpaired particle
density $n_{\uparrow} - n_{\uparrow \downarrow}$.}
\label{fig:te} 
\end{figure}

The dynamics after releasing the particles from the potential show
a striking two-fluid behaviour. Pairs (doublons) and
excess unpaired majority particles expand
as effectively non-interacting fluids, as can be seen from Figure \ref{fig:te}.
Indeed, we have compared the dynamics with strong interaction between
the spins to the dynamics of non-interacting particles and verified
that the doublons and unpaired particles in the strongly interacting
limit expand {\it qualitatively} just like non-interacting
particles would (see supplementary information Figures 2-5 for the
$U=0$ results \cite{Kajala:8BtvNK09}).  The important difference is,
however, that {\it the velocity of the expansion is changed} with
respect to the non-interacting case. The doublons expand with
velocities up to $\frac{4J^2}{U} \sin(k_{F \downarrow})$.  And what is
crucial for our proposal for observing the FFLO state: the unpaired
particles expand with velocities up to $2J \sin (q)$, where $q$ is the
FFLO momentum. In the $U \gg J$ case shown here the 
unpaired particle velocity is larger than the pair one, but in general 
the relative velocities of pairs vs. unpaired particles is not essential 
for the method since the up and down components can be imaged 
separately, see e.g. \cite{Liao:2010bu}.

\begin{figure}
\includegraphics[width=0.4\textwidth]{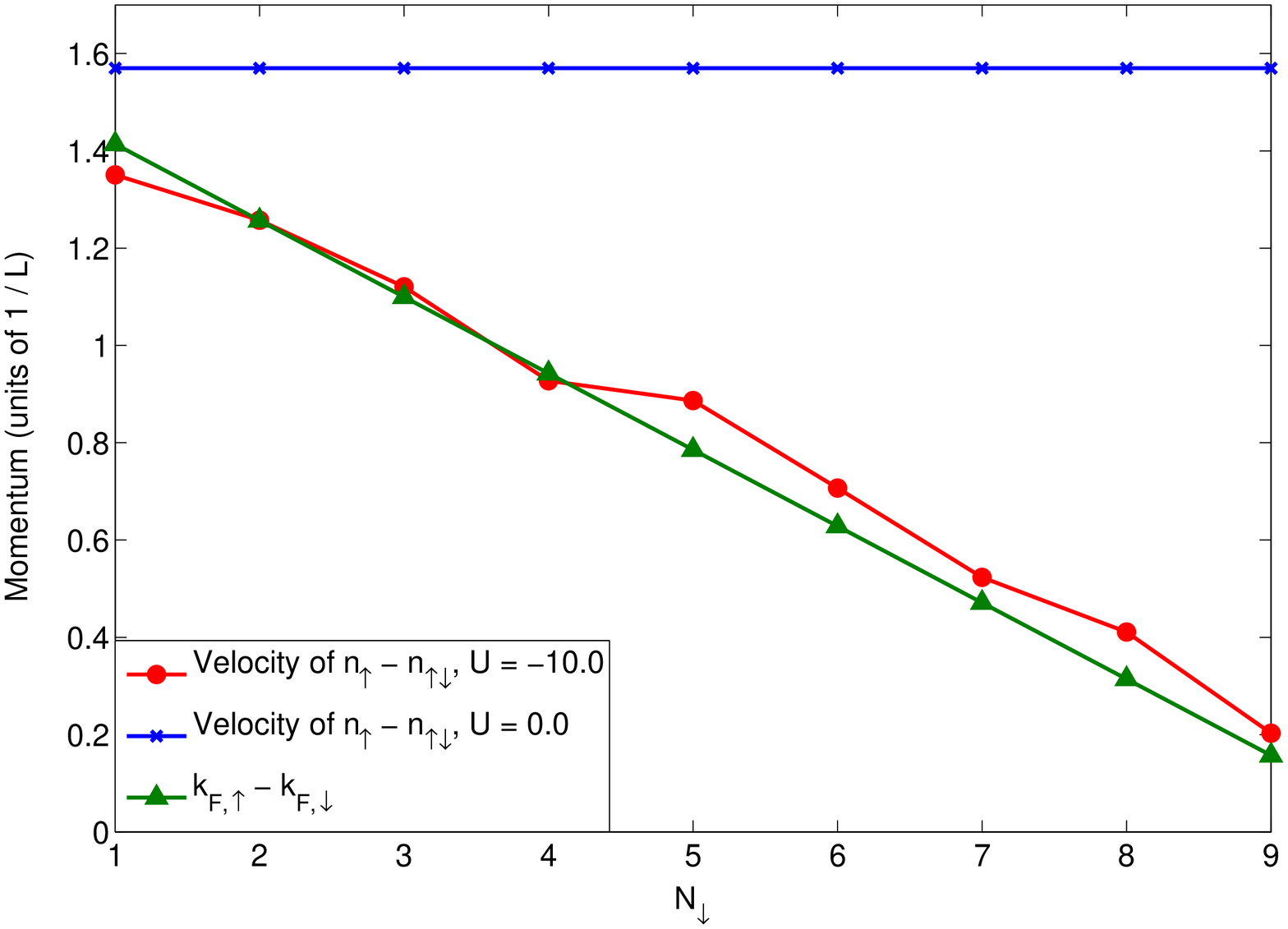} 
\includegraphics[width=0.4\textwidth]{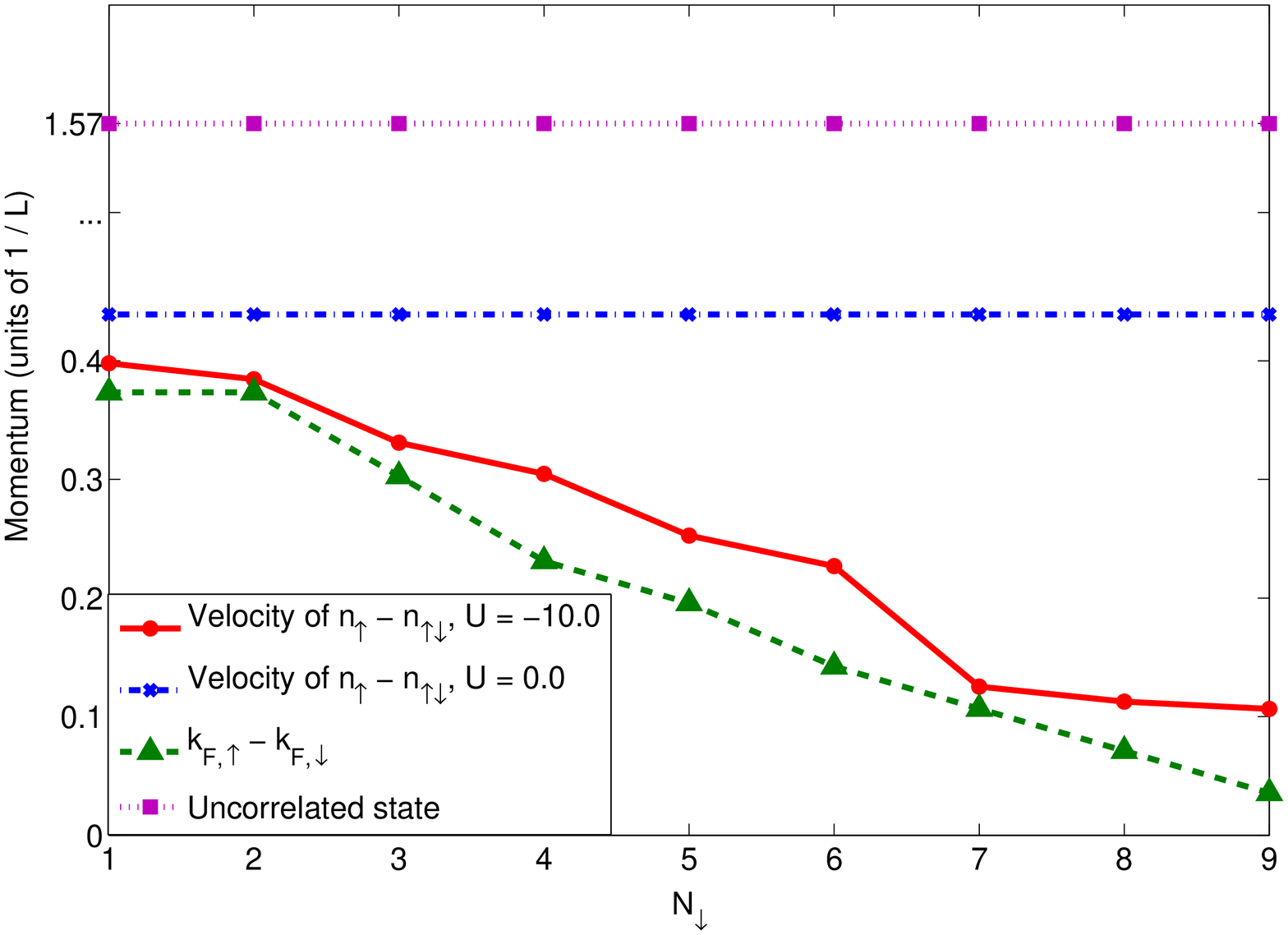} 
\caption {(Color online) \textbf{a}: The FFLO momentum $q$ determined from the
  edge expansion velocity of the unpaired cloud, compared to the $q$ of the
  ground state, as a function of $N_\downarrow$ describing the
imbalance ($N_\uparrow=10$). We also show the expansion momentum $k \neq q$
  obtained in the case of a non-interacting gas, and a non-FFLO state
  without pair coherence. \textbf{b}: Same as \textbf{a},
  but having initially a shallow harmonic trap ($V_{ho}=0.0003$, $C=75.5$).}
\label{fig:MAIN_RESULT} 
\end{figure}

A wavefront corresponding to $q$ clearly separates
from the rest of the unpaired particle cloud during initial dynamics
($t = 0 - 10 \frac{1}{J}$), after which it moves with a constant
velocity at the edge of the cloud, see
Fig.\ref{fig:te}$_{\textbf{b}}$. Therefore, by measuring the expansion
velocity of the cloud edge ($v_{exp}$), e.g. from the maximum gradient
of the density (see supplementary information Figure 1 \cite{Kajala:8BtvNK09}),
one obtains the FFLO momentum from
\begin{equation}
\label{FFLO_q}
q = \arcsin\left(\frac{v_{exp}}{2J}\right).
\end{equation}
The momentum $q$ obtained in this way from our simulations is compared
in Figure \ref{fig:MAIN_RESULT}$_{\textbf{a}}$ to the FFLO momentum as
given by the definition $q = k_{F, \uparrow} - k_{F, \downarrow}$ (and
by the peak in the momentum pair correlation function of the ground state,
c.f.\ Figure \ref{fig:gs}$_{\textbf{b}}$).  For reference, we show
the expansion velocities in the cases of a non-interacting gas, and a
non-FFLO state without pair coherence (discussed later in the text).
Only in the case of the FFLO state does the expansion velocity
depend on the imbalance. The $q$ extracted from the simulations via
Eq.(\ref{FFLO_q}) matches excellently the expected FFLO momentum.

\begin{figure}
\includegraphics[width=0.45\textwidth]{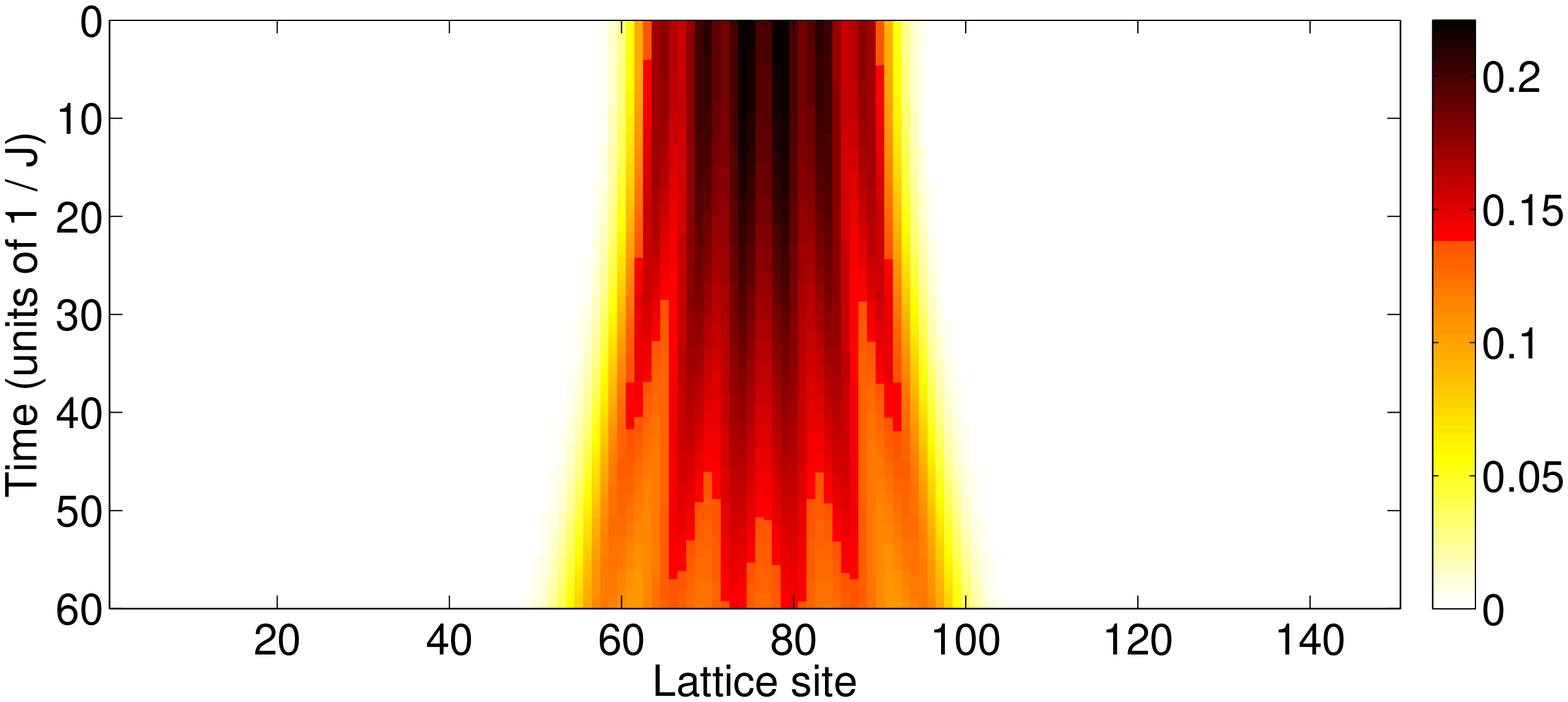} 
\includegraphics[width=0.45\textwidth]{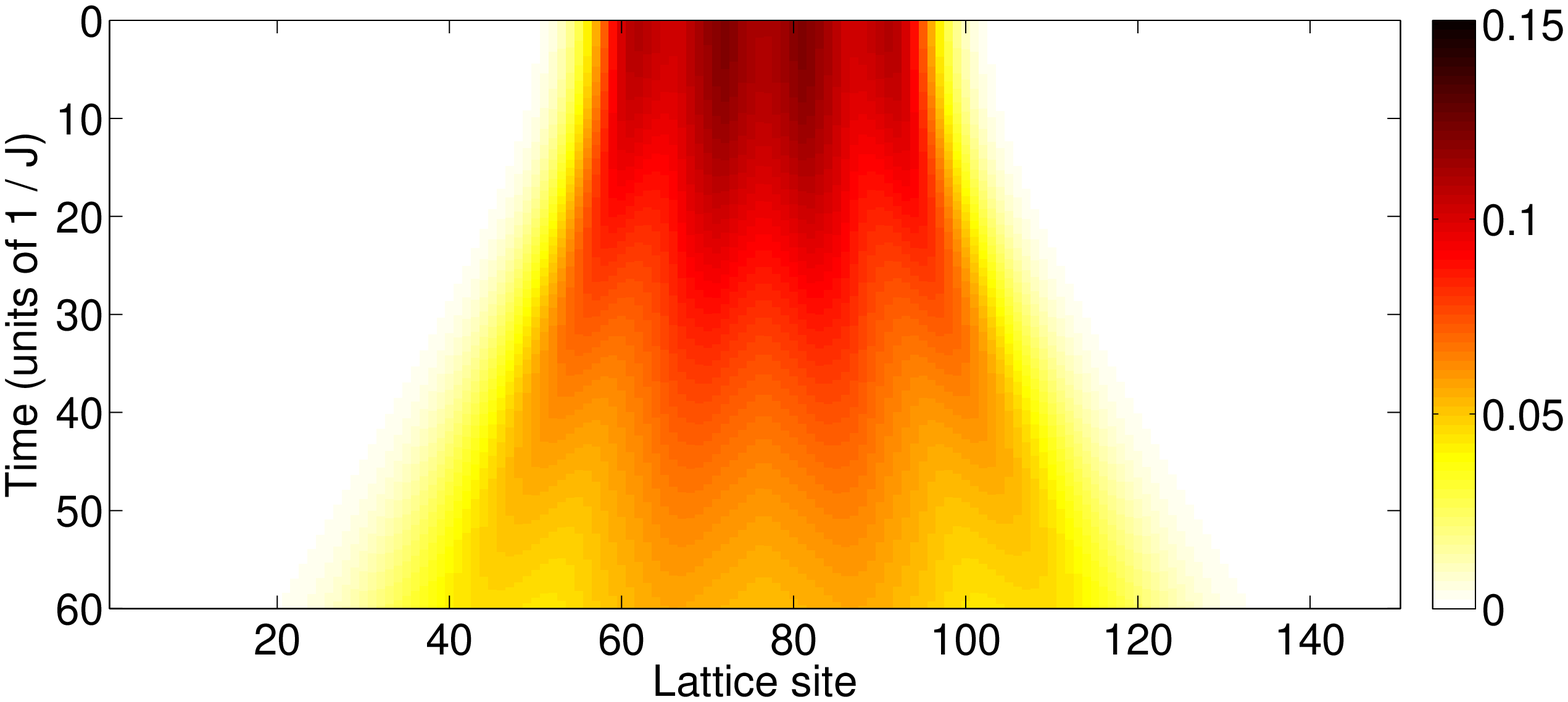} 
\caption {(Color online) \textbf{a}: The time development of the doublon density
  $n_{\uparrow \downarrow}$ with the same parameters as in Figure
  \ref{fig:gs} but the initial trap being harmonic ($V_{ho}=0.0003$, $C=75.5$) instead of a box
  trap.  \textbf{b}: The time development of the
  unpaired particle density $n_{\uparrow} - n_{\uparrow \downarrow}$.}
\label{fig:harmonic} 
\end{figure}

In the experimentally relevant case of a harmonic trap, $q$ can be
determined in the same way as discussed above.  Figure
\ref{fig:MAIN_RESULT}$_{\textbf{b}}$ shows the comparison of $q$
obtained from the edge expansion velocity to the FFLO momentum in a
harmonic trap.  The time evolution is shown in Figure \ref{fig:harmonic}. 
In $q=k_{F\uparrow} - k_{F\downarrow}$, 
the Fermi momenta are the highest momenta
in the harmonic oscillator eigenstates of the quantum numbers 
$n_{F\uparrow}$ and
$n_{F\downarrow}$, respectively (see supplementary information
\cite{Kajala:8BtvNK09}).  In the limit of a shallow trap and large particle
number, one approaches the box potential case and 
a local density approximation (LDA)
argument can be used for the extraction of $k_{F\,\sigma}$ \cite{Tezuka:2008fp}. 

The interpretation of our numerical results in terms of the expansion
of two fluids is supported by a rigorous Bethe-ansatz analysis of the
one-dimensional Fermi-Hubbard Hamiltonian
\cite{Giamarchi:2004vu,Essler:2005uw,Zhao:2010hk}. In the
strong-coupling limit, the system can be described as two weakly
interacting spinless Fermi gases, corresponding to pairs and unpaired
particles whose maximum group velocities are the respective Fermi
velocities. We thus expect the time evolution of the system to be
approximately generated by the free-particle Hamiltonians for unpaired
particles and pairs. In the case of a lattice model, the group
velocity for each momentum component is $v_k=2 \tilde{J}
\sin\left[k\right]$, where the constant $\tilde{J}$ depends on the
nature of the particle considered. In our pair/unpaired particle
2-fluid model stemming from the Hubbard Hamiltonian, we have
$\tilde{J}=J$ for the unpaired particles and $\tilde{J}=2J^2/U$ for
the pairs.  For $k<\pi/2$, the relation $v_k=2 \tilde{J}
\sin\left[k\right]$ thus allows to establish a connection between the
maximal expansion velocity and the Fermi momentum of each component
(note that for $k \geq \pi/2$ the maximal expansion velocity is given
by $2J$). 

Based on our numerical findings, we now assume that the
unpaired-particle Fermi momentum and the FFLO momentum share the same
value $q=k_{F\,\uparrow}-k_{F\,\downarrow}$. This is supported by
Bethe-ansatz analysis in the continuum case \cite{Oelkers:2006hi} and
is also intuitive: in the limit of strong interactions the pair
kinetic energies are small and it is energetically favourable that the
ground state structure allows the lowest momentum states up to $q$ to
be mainly occupied by unpaired majority particles.  Thus measuring the
maximal expansion velocity of the unpaired particles gives access to
the value of the FFLO momentum. As seen in Figure
\ref{fig:MAIN_RESULT}, this scenario agrees with the numerical
results, providing a clear signature of the FFLO state. The lattice
reproduces the continuum-case dynamics for low densities
\cite{Yang:2001iz,Zhao:2010hk} (we have tested also the low density
limit), and our proposal is thus expected to be suitable for the
detection of the FFLO state in experiments of the type
\cite{Liao:2010bu}.  Moreover, the proposed method should be robust
with respect to averaging over an array of 1D tubes, c.f. \cite{Liao:2010bu}.

Our simulations describe the zero temperature case. Quantum Monte
Carlo calculations suggest that there is a finite-temperature
precursor for the FFLO state which involves pairing but no FFLO-type
correlations \cite{Wolak:2010ib}. Is the signature we propose for the
FFLO state distinguishable from any traces of a state with pairing but
no FFLO correlations?  Simulating exact dynamics at finite temperature
is a formidable task and well beyond the state-of-the-art.  However,
to consider pairing without FFLO correlations, we have simulated
the dynamics of various initial states which have the same
average densities of unpaired and paired particles as the
corresponding FFLO states, but which possess no correlations between
the particles or between the pairs.  
Such an uncorrelated state can be written as an ensemble of product states, 
in which each of the product states have completely spatially 
localized pairs and single majority particles.
Choosing the constituents of the ensemble randomly and giving them equal weight,
we obtain the results shown in Figure \ref{fig:MAIN_RESULT}: the
expansion velocity does not depend on the polarization. It is simply
given by $2J$ and $4J^2/U$ for unpaired particles and pairs,
respectively.  Therefore, observing a change of the unpaired particle
expansion velocity with the polarization, such that it has the
functional dependence $q=k_{F\,\uparrow}-k_{F\,\downarrow}$, is a
genuine signature of the FFLO correlations, and cannot be achieved for
a non-correlated, paired state with the same imbalance. When lowering
the temperature, emergence of a $q$-dependent expansion front on top
of a thermal state background would reveal that the FFLO state has
been reached.

To conclude, we have shown that like in many classic ultracold gas
experiments, also in the case of the long-sought-for FFLO state, the
expansion of the cloud gives an exceedingly simple way of determining
the nature of the initial state.  Our exact quantum many-body
simulations of the 1D imbalanced gas present a clear two-fluid
behaviour with characteristic expansion velocities. We show that the
matching of the expansion velocity of the unpaired majority cloud with
the expected FFLO momentum provides an unambiguous signature of the
FFLO state.  

Acknowledgements: We thank R.G. Hulet for useful discussions.  This work
was supported by the Academy of Finland (Projects 213362, 217043,
217045, 210953, 135000, and 141039), ERC (Grant
No. 240362-Heattronics), and conducted as a part of a EURYI scheme
grant (see www.esf.org/euryi).  Computing resources were provided by
CSC - the Finnish IT Centre for Science.

\section{Supplementary Material}

\section{Dynamics of the Gaudin-Yang model in the strongly attractive regime}
\label{sec:GY}

As shown in \cite{Oelkers:2006hi:s}, the Bethe-ansatz solution for the
Gaudin-Yang model in the strongly interacting regime can be written in
terms of the quasi-momenta $k_i$ ($n$ for unpaired particles, $p$ for
pairs) and the spin roots $\Lambda_p$
\begin{align}
  \label{eq:BAsol1}
  & k_n= \frac{\pi}{L} n + \frac{\delta n}{c L} & 
       n=\left\{1 \dots N_\uparrow-N_\downarrow
       \right\} \\
   \label{eq:BAsol2}
  &\Lambda_p = \frac{\pi}{L} p + \frac{\epsilon p}{c L} &
       p=\left\{1/2,1,3/2 \dots N_\downarrow/2\right\}  \\
  \label{eq:BAsol3}
   &k_{p\,\pm}= \Lambda_p \pm \frac{1}{2} i c 
\end{align}
where $L$ represents the size of the system and $2c$ the contact
interaction strength, while $\delta$ and $\epsilon$ are constants
related to $N_\uparrow/L$ and $N_\downarrow/L$. Note that $c$ has been
rescaled so that the physical value of the interaction strength is
given by $ \tilde{c} = \hbar^2 c
/2m$. Eqs. (\ref{eq:BAsol1}-\ref{eq:BAsol3}) allow us to rewrite the
dispersion relation for the system
\begin{equation}
  \label{eq:disp1}
  E= \frac{\hbar}{2m} \left( \sum_p k_p^2 + \sum_n k_n^2\right) 
\end{equation}
in the following form 
\begin{align}
  \label{eq:dispRel} 
    E= \frac{\hbar^2}{2m}\left[\sum_p \left( 2 \Lambda_p^2 -
    c^2/2\right) + \sum_n k_n^2\right].
\end{align}
From Eq. (\ref{eq:dispRel}), we recognize the pair contribution to the
kinetic energy ($2 \Lambda_p^2$), along with the pair-binding energy
($c^2/2$) and the single-particle kinetic energy ($k_n^2$). We can
thus state that since, in the strongly interacting limit, the binding
energy depends on the interaction energy $c$ only, it is independent
of the Fermi energy. Remembering the definition of $\Lambda_p$, for
large interaction (i.e. $1/c \to 0$ ), it is possible to rewrite the
pair kinetic energy term as
\begin{equation}
  \label{eq:Epair}
   E_{s\,{\rm kin}}=\frac{\hbar}{m} \left(\frac{\pi}{2L} s \right)^2
\end{equation}
with $s= \left\{1,2 \dots N_\downarrow\right\}$. The term
$\frac{\pi}{L} s$ can be identified with the quasi momentum
$\kappa_{s}$ of a free particle in a system of size $L$. If we
calculate the velocity of such object
\begin{equation}
  \label{eq:vPair}
  v_{s}=\frac{\partial E_{s\,{\rm kin}}}{\partial \hbar \kappa_{s}}=
\frac{\hbar \kappa_{s} }{2m},
\end{equation}
we can conclude that, in the strong-coupling limit, along with
$N_\uparrow-N_\downarrow$ excitations with group velocity $\hbar
k_n/m$ (particles), $N_\downarrow$ excitations moving with group
velocity $\hbar \kappa_s/2m$ are present. We identify these
excitations with pairs with binding energy $c^2/2$ 
and mass $2m$.


\section{Uncorrelated State Dynamics}

The uncorrelated state dynamics originates from the dynamics of 
localized particles released into 
the lattice, similiar to the case previously studied in \cite{Kajala:2011ho}.
The expansion wavefronts which emerge have velocities $2J$ and $\frac{4J^2}{U}$,
which is in accordance with our understanding that we see 
the maximum velocities of unpaired particles and pairs in the dynamics.
Note that localization to a lattice site corresponds in the Fourier space
to employing all momenta in the band.  
The maximum expansion velocities observed correspond to momenta
$k = \frac{\pi}{2}$ since, due to the lattice dispersion $E = - 2 \tilde{J} \cos(k)$,  
$v_{exp} = \frac{dE}{dk} = 2 \tilde{J} \sin(k)$, where we have
$\tilde{J} = J$ for unpaired particles and $\tilde{J} = \frac{2J^2}{U}$ 
for pairs.

\section{Determining the edge expansion velocity}

\begin{widetext}

\begin{figure}
\includegraphics[width=0.99\textwidth]{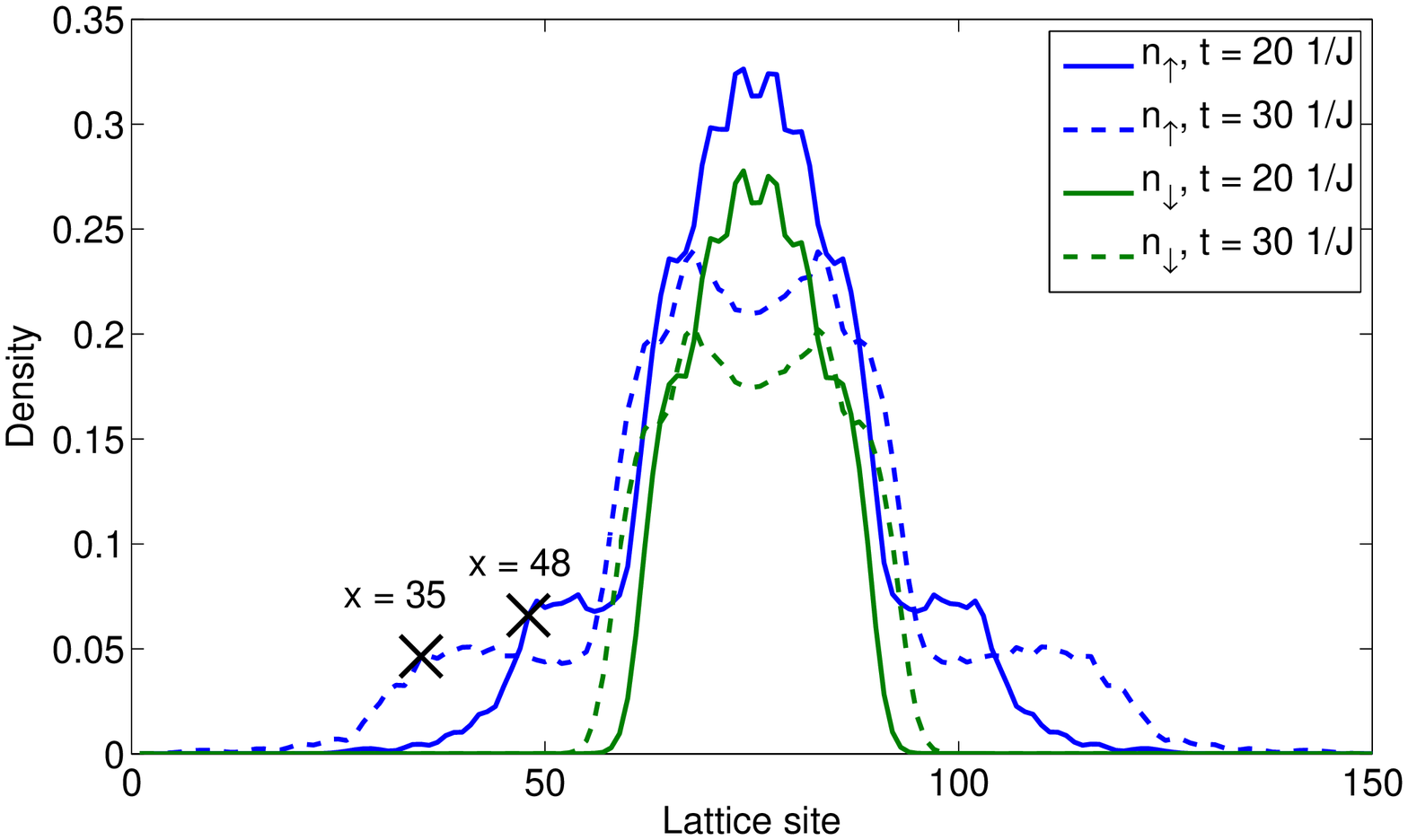} 
\includegraphics[width=0.99\textwidth]{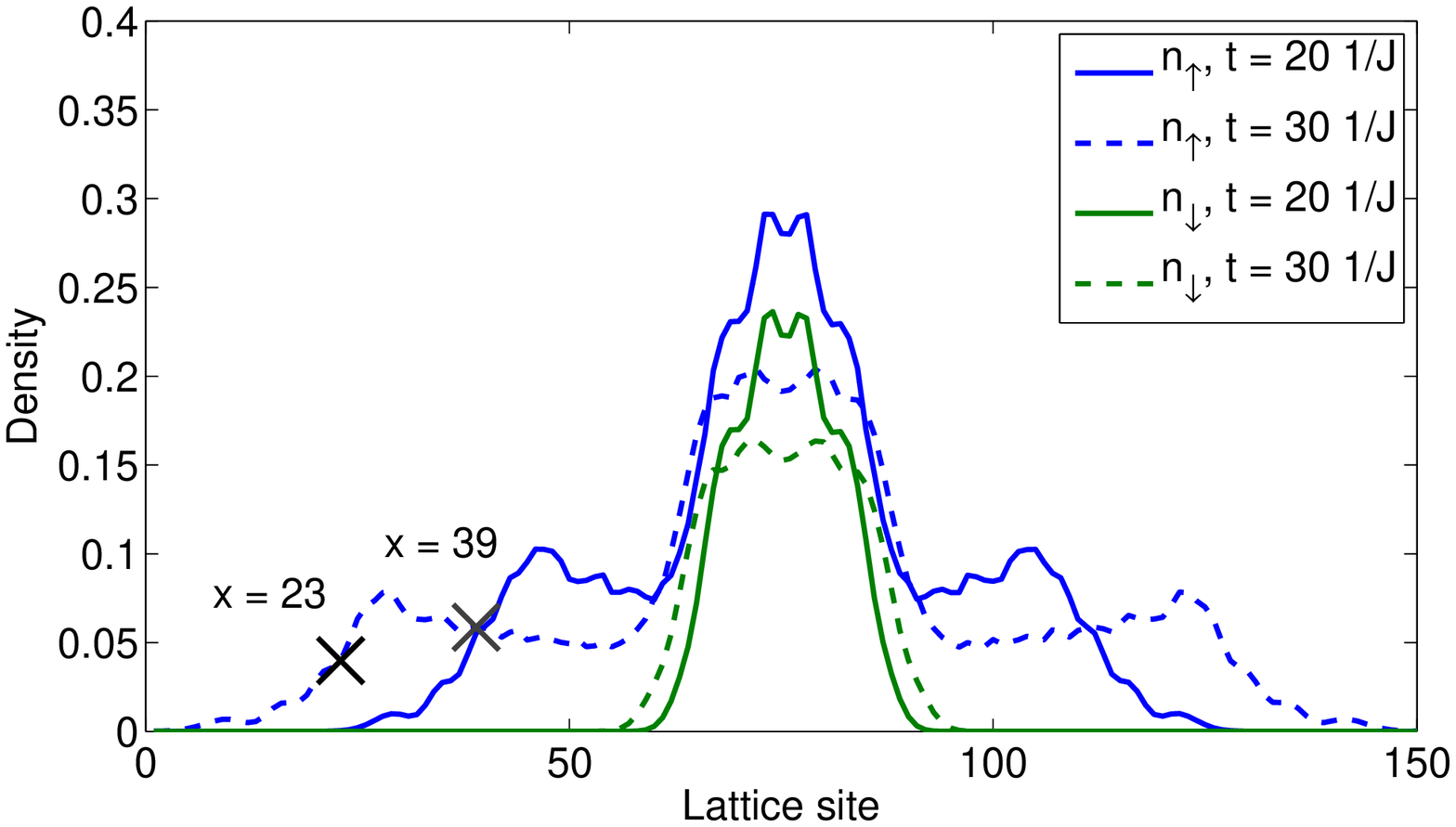} 
\caption {\textbf{a}: Obtaining the expansion velocity from the position
of the maximum gradient of unpaired particle density at the edge. The FFLO wavevector 
is obtained from $q = \arcsin(\frac{v_{exp}}{2J})$. Substituting gives
$q = \arcsin(\frac{\frac{13 J}{10}}{2J}) \frac{1}{L} = 0.71 \frac{1}{L}$. 
Particle numbers here are $N_{\uparrow} = 10, N_{\downarrow} = 6$, and initially 
the gas was confined in a box potential. 
\textbf{b}: The same as \textbf{a} but for $N_{\uparrow} = 10, N_{\downarrow} = 4$. The
FFLO wavevector in this case is 
$q =  \arcsin(\frac{\frac{16 J}{10}}{2J}) \frac{1}{L} = 0.93 \frac{1}{L}$.} 
\label{fig:gradient} 
\end{figure} 

\end{widetext}

The edge expansion velocity has been determined using the numerically obtained 
density profiles at different times after the release from 
the trapping potential. We are interested in the maximum group
velocity of the unpaired component, as this gives the FFLO momentum $q$. 
In the simulations it is seen that this unpaired wavefront 
separates from the rest after initial dynamics ($t = 1 - 10 \frac{1}{J}$) 
after which it travels at constant velocity at the edge of the cloud. 
 The velocity of the edge wavefront has been determined by calculating
the change of the position of the maximum gradient of the unpaired majority particle
density at the edge. Figure \ref{fig:gradient} illustrates how this 
has been done. 

\section{Comparing non-interacting and strongly interacting expansion dynamics}

\begin{figure}[!H]
\includegraphics[width=0.4\textwidth]{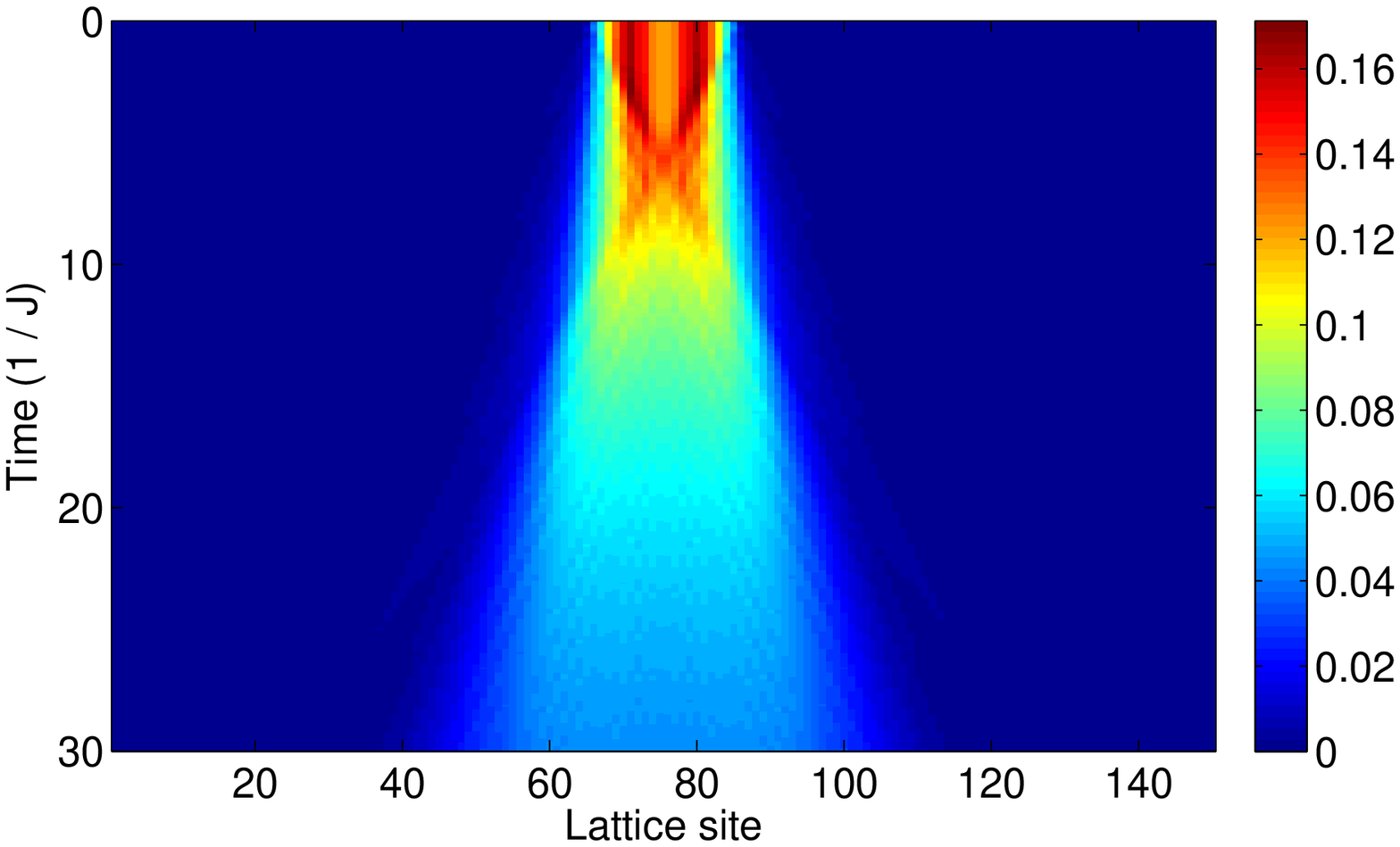} 
\includegraphics[width=0.4\textwidth]{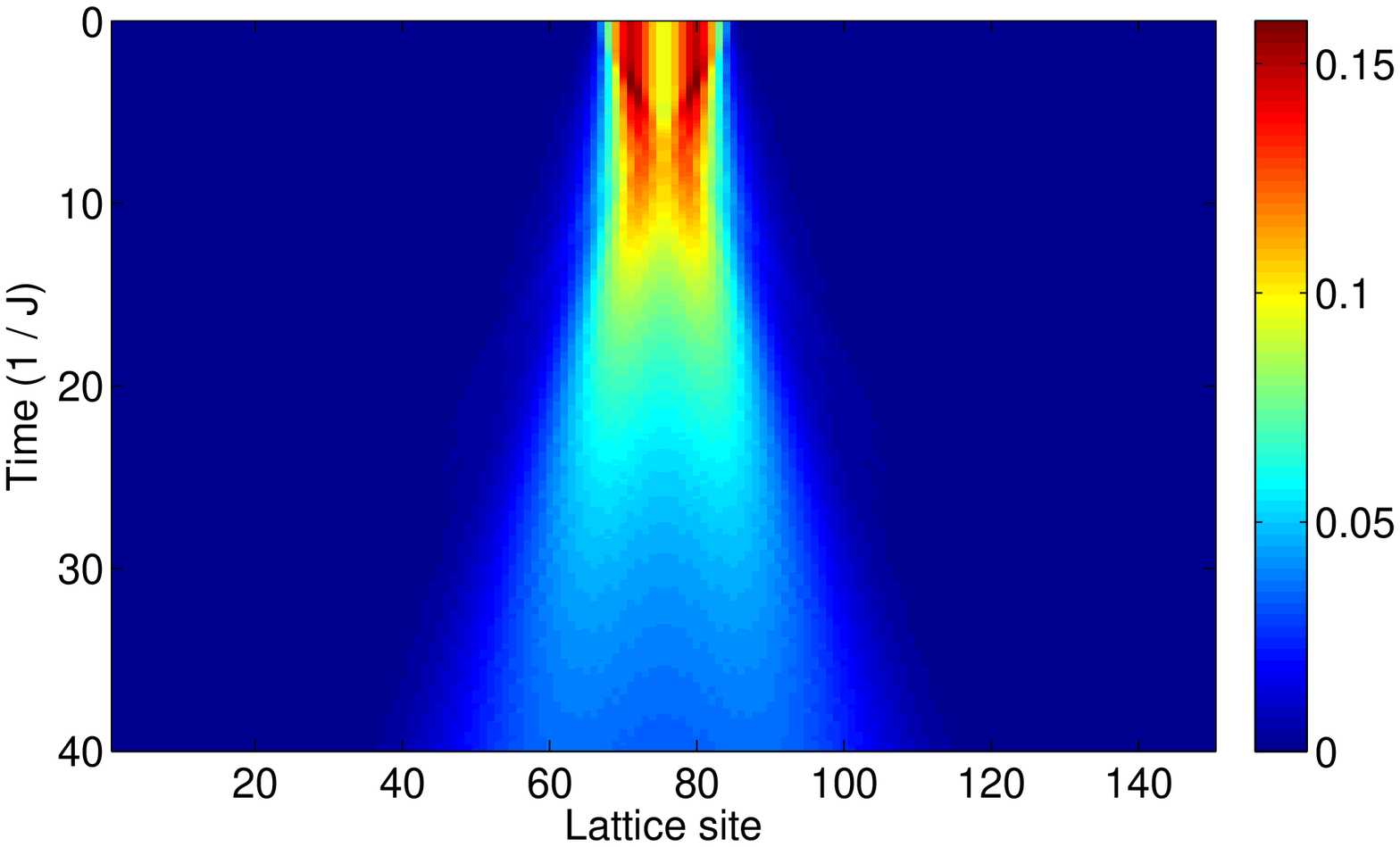} 
\caption {\textbf{a}: The density profile of 
unpaired particles $n_{\uparrow}(t) - n_{\uparrow\downarrow}(t)$ after the release of 
the gas from 
the box potential when $N_{\uparrow} = 10$, $N_{\downarrow} = 8$, $U = -10.0 J$.
\textbf{b}: The density profile $n_{\uparrow}(t)$ of two noninteracting 
particles released from the box potential ($N_{\uparrow} = 2$, $N_{\downarrow} = 0$).
Notice that the time axis is different in the two plots, showing how the
expansion is qualitatively similiar but occurs at a different velocity.} 
\label{fig:comp1} 
\end{figure}

\begin{figure}[!H]
\includegraphics[width=0.4\textwidth]{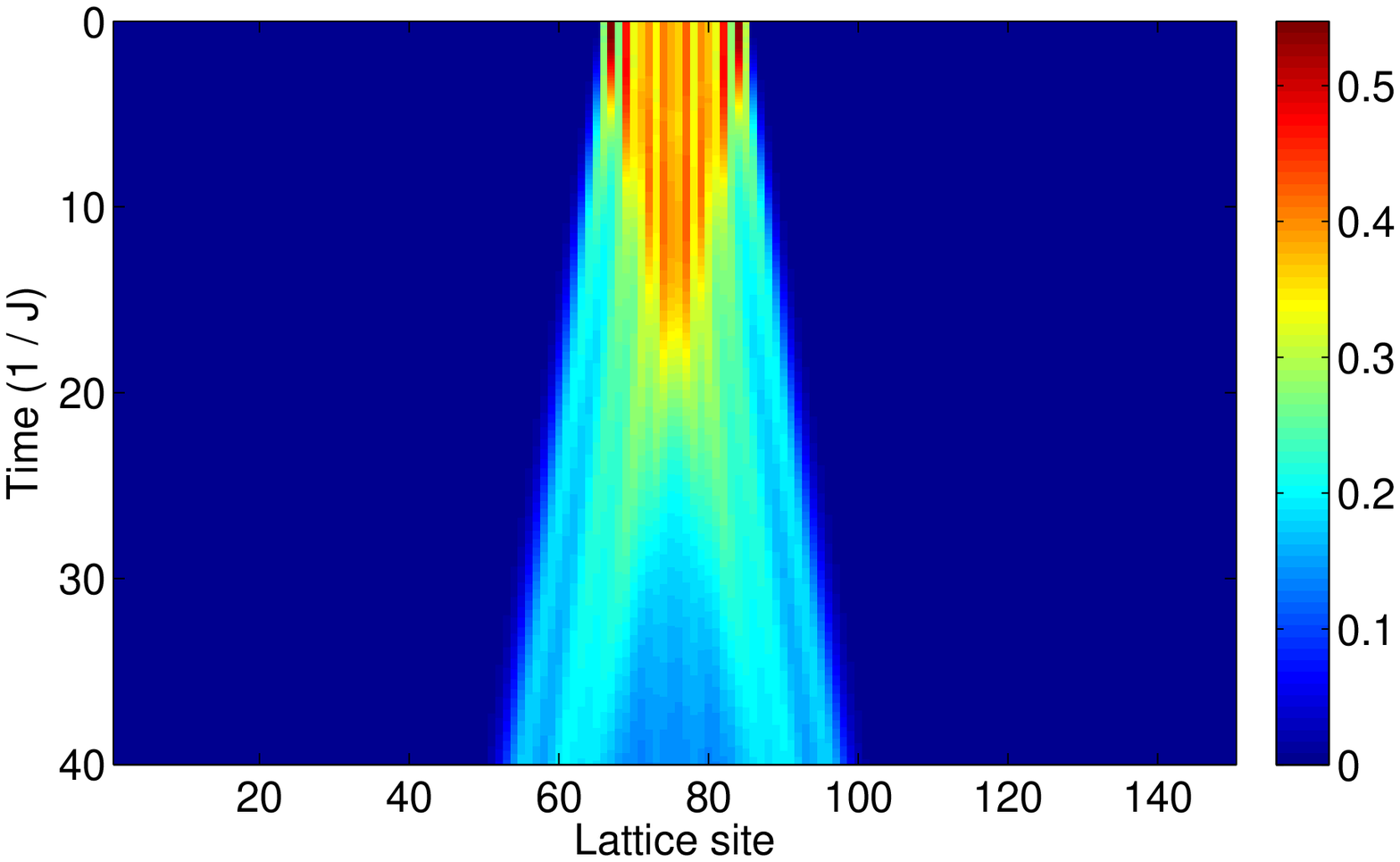} 
\includegraphics[width=0.4\textwidth]{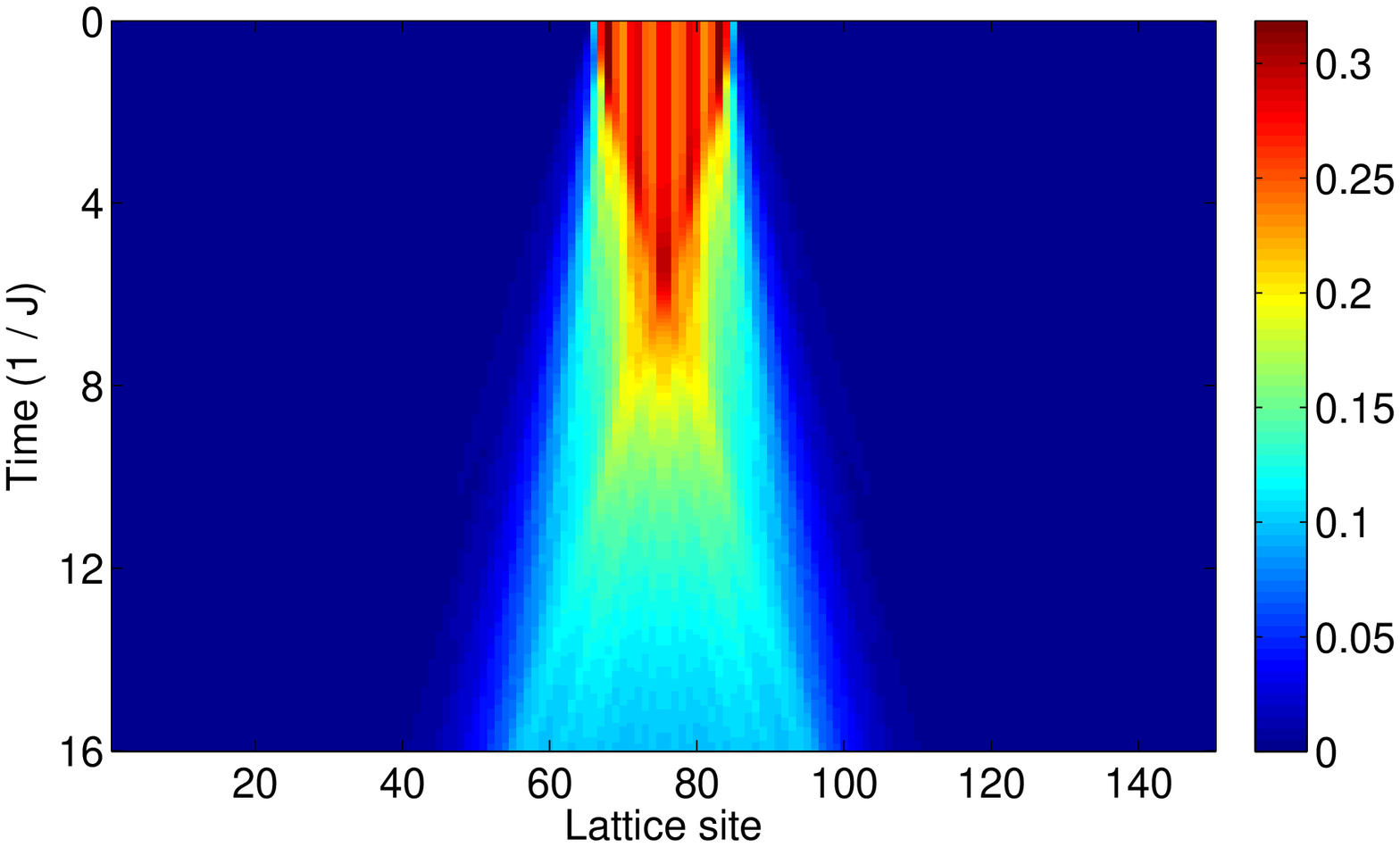} 
\caption {\textbf{a}: The density profile of paired particles 
$n_{\uparrow\downarrow}(t)$ after the release of the gas from 
the box potential when $N_{\uparrow} = 10$, $N_{\downarrow} = 8$, and $U = -10.0 J$.
\textbf{b}: The density profile $n_{\uparrow}(t)$ of eight noninteracting 
particles released from the box potential ($N_{\uparrow} = 8$, $N_{\downarrow} = 0$).
Notice again the scaling of the time axis.} 
\label{fig:comp2} 
\end{figure}

\begin{figure}[!H]
\includegraphics[width=0.4\textwidth]{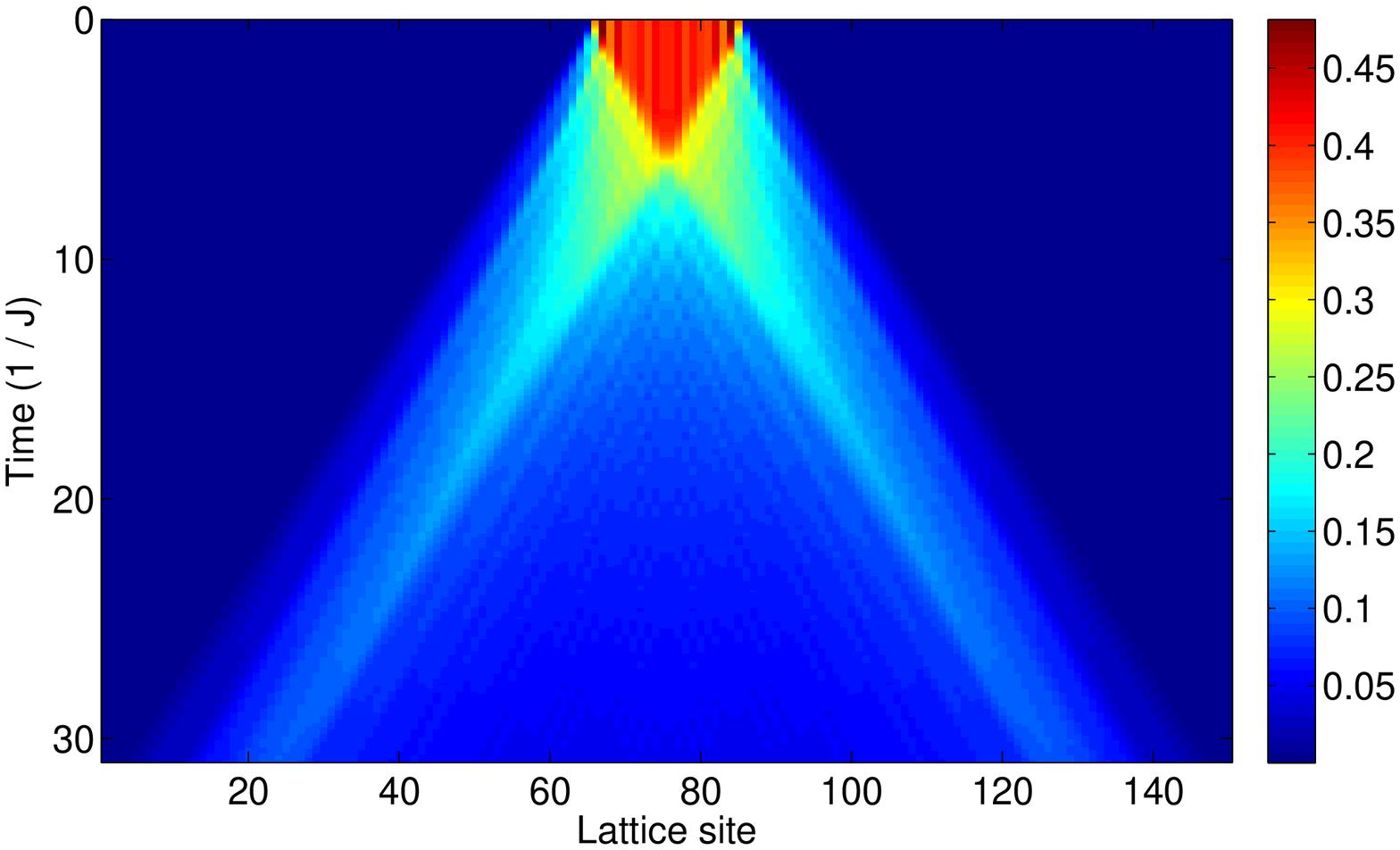} 
\includegraphics[width=0.4\textwidth]{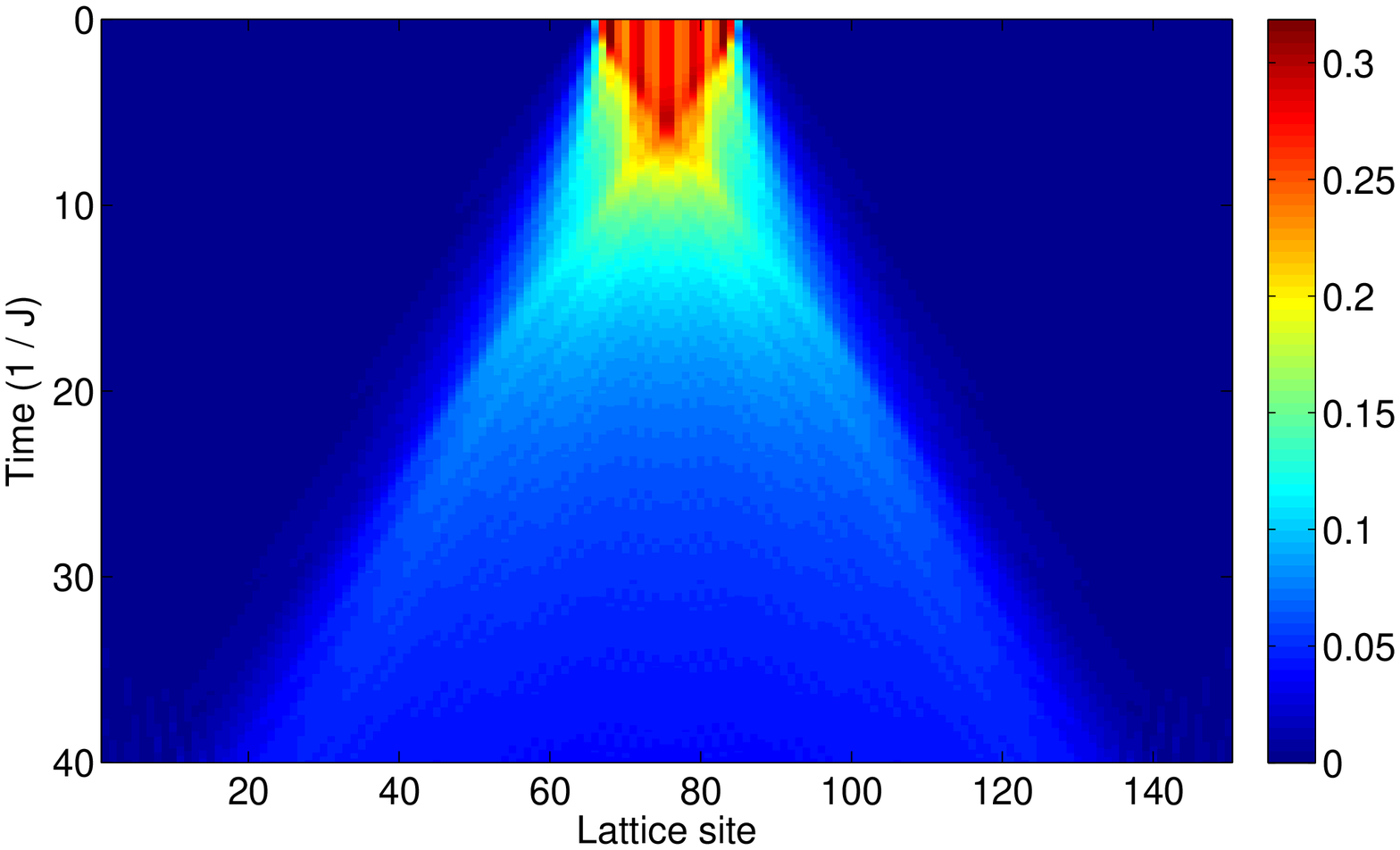} 
\caption {\textbf{a}: The density profile
$n_{\uparrow}(t) - n_{\uparrow\downarrow}(t)$ for parameters
$N_{\uparrow} = 10$, $N_{\downarrow} = 2$, and $U = -10.0 J$.
\textbf{b}: The density profile $n_{\uparrow}(t)$ for 
$N_{\uparrow} = 8$ and $N_{\downarrow} = 0$.} 
\label{fig:comp3} 
\end{figure}

\begin{figure}[!H]
\includegraphics[width=0.4\textwidth]{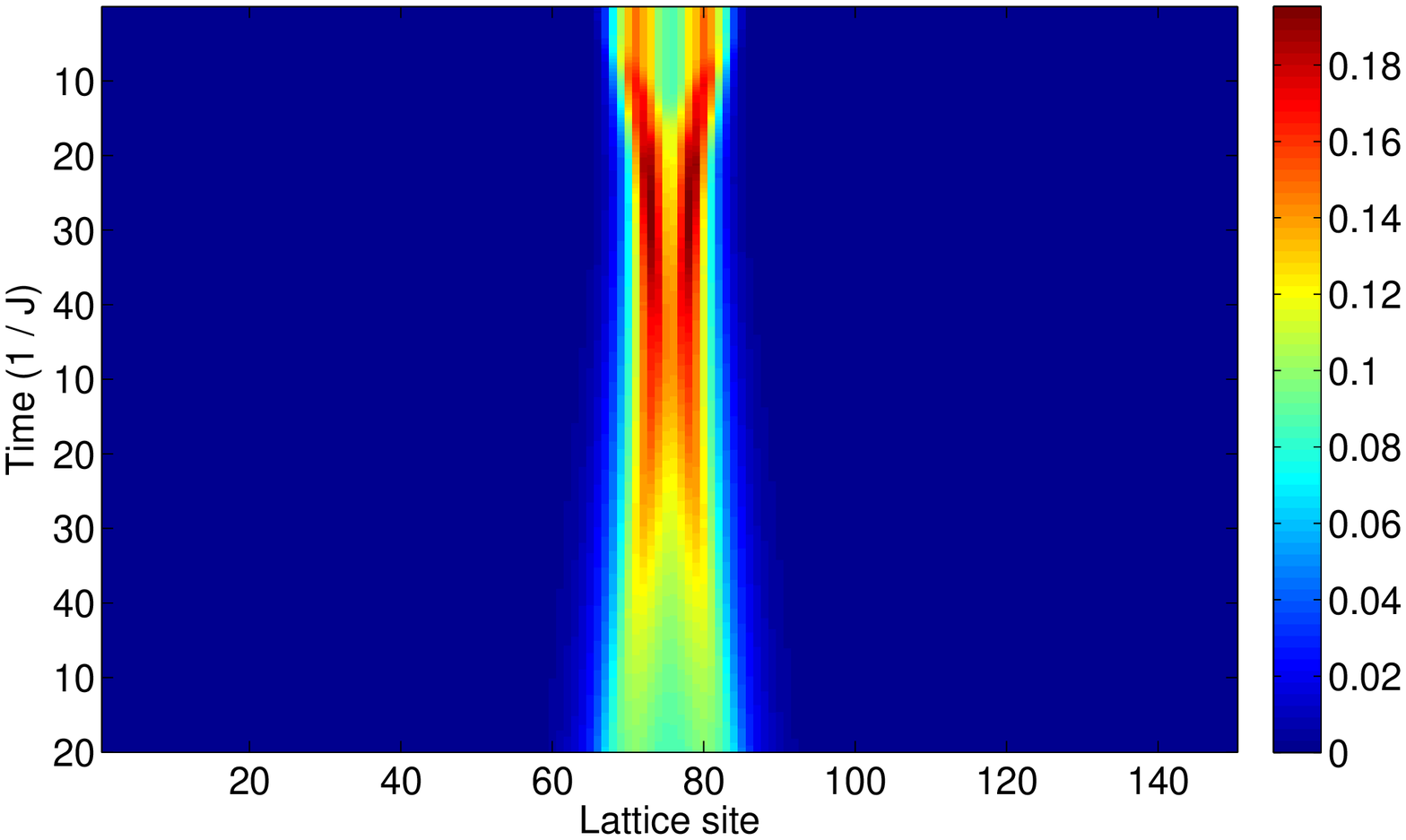} 
\includegraphics[width=0.4\textwidth]{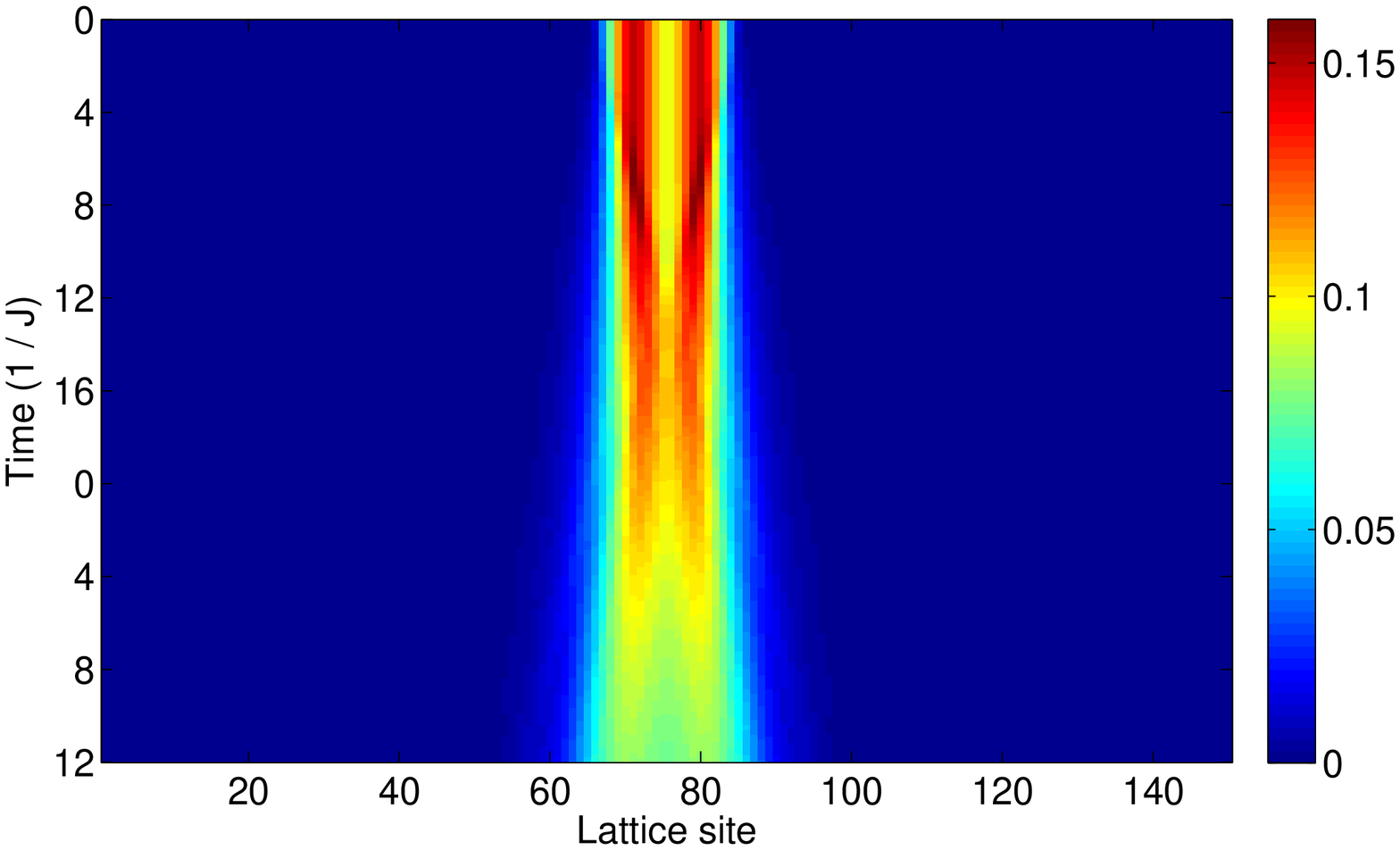} 
\caption {\textbf{a}: The density profile $n_{\uparrow\downarrow}(t)$ for 
$N_{\uparrow} = 10$, $N_{\downarrow} = 2$, and $U = -10.0 J$.
\textbf{b}: The density profile
$n_{\uparrow}(t)$ for $N_{\uparrow} = 2$ and $N_{\downarrow} = 0$.} 
\label{fig:comp4} 
\end{figure}

Figures \ref{fig:comp1} - \ref{fig:comp4} illustrate how the 
expansion profiles of X doublons and Y unpaired particles look qualitatively 
like the expansion profiles of X and Y non-interacting particles, respectively, 
but have different velocities compared to the non-interacting case.
The velocity in the strongly interacting regime is $2J\sin(q)$ for the
unpaired particles and $\frac{4J^2}{U}sin(k_{F \downarrow})$ for the pairs 
(down is the minority species). In the noninteracting case, the expansion
velocity of species $\sigma$ is given by $2J\sin(k_{F \sigma})$. 

\section{Additional data}

\begin{figure}[p]
\includegraphics[width=0.4\textwidth]{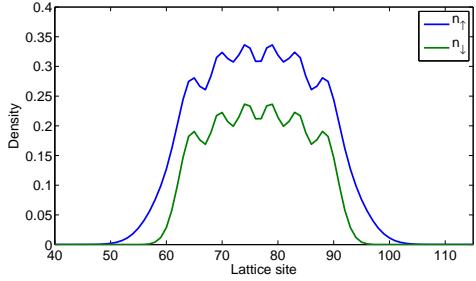} 
\caption {The up and down particle densities $n_{\uparrow}$ and $n_{\downarrow}$ for 
the ground state with harmonic trapping and parameters
$N_{\uparrow} = 10$, $N_{\downarrow} = 6$, $L = 150$, 
$U = -10.0 J$, and $V_{ho} = 0.0003 J$.} 
\label{fig:gs_trap} 
\end{figure}

\begin{figure}[!H]
\includegraphics[width=0.425\textwidth]{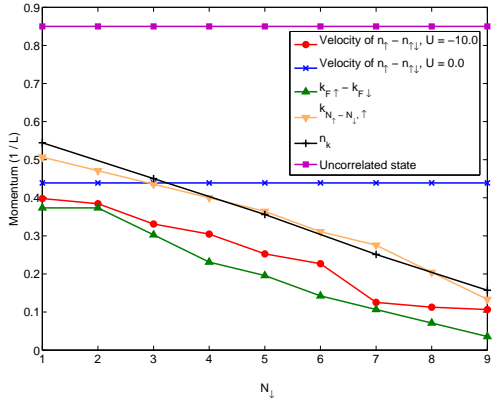} 
\caption {The FFLO momentum $q$ in a harmonic trap
as determined from the edge expansion velocity,
the momentum of the edge expansion for the noninteracting case, 
$k_{F \uparrow} - k_{F \downarrow}$ as obtained from 
the peaks in momentum distribution of highest occupied 
(noninteracting) harmonic oscillator eigenstates (see text for explanation),  
$k_{(N_\uparrow-N_\downarrow)\uparrow}$, position of the peaks in the pair momentum correlation $n_k$, 
and the momentum of the edge expansion given by an uncorrelated state.}
\label{fig:q_trap_all} 
\end{figure}

Figure \ref{fig:gs_trap} 
depicts the ground state of the 1D FFLO superfluid in a harmonic
trap as described in the main text. 
The density is low enough so that the lattice 
result corresponds to the continuum case in the strongly interacting regime. 
Corresponding to this ground state, Figure \ref{fig:q_trap_all} shows the 
momenta involved in FFLO pairing in the 
case of a harmonic trap with small particle number. 

\begin{figure}[!H]
\includegraphics[width=0.425\textwidth]{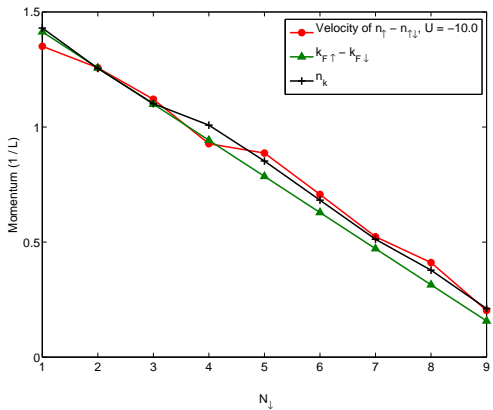} 
\caption {The FFLO momentum $q$ in a box potential
as determined from the edge expansion velocity,
$k_{F \uparrow} - k_{F \downarrow}$ as obtained from 
the highest occupied noninteracting box eigenstates 
($k_{F \sigma} = \frac{N_{\sigma}\pi}{L}$),
and the position of the peaks in the pair momentum correlation $n_k$.}
\label{fig:q_box_all} 
\end{figure}

\begin{figure}[!H]
\includegraphics[width=0.4\textwidth]{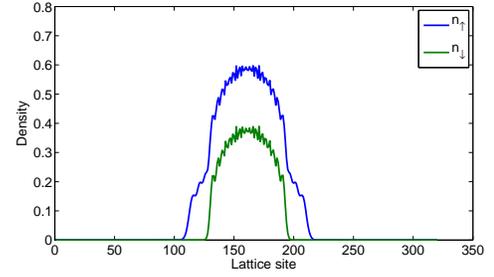} 
\includegraphics[width=0.4\textwidth]{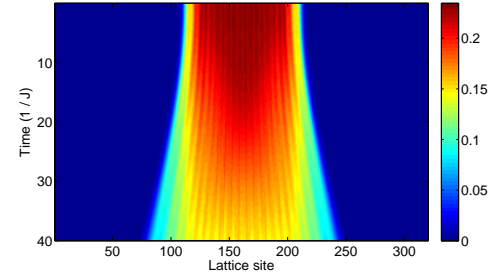} 
\includegraphics[width=0.4\textwidth]{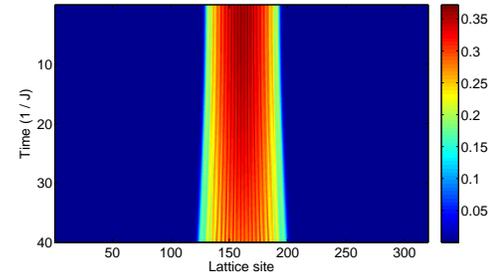} 
\includegraphics[width=0.4\textwidth]{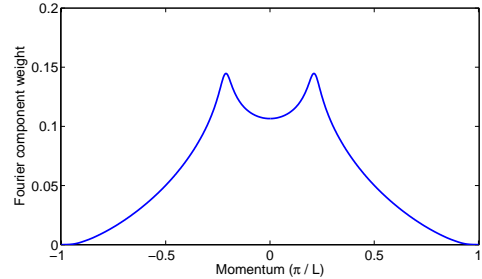} 
\caption {\textbf{a}: The ground state in a harmonic trap with the parameters
$L = 320$, $N_{\uparrow} = 40$, 
$N_{\downarrow} = 20$, $V_{ho} = 0.0003$, and $U = -10.0$. \textbf{b}: The corresponding
time evolution of unpaired particles $n_{\uparrow}(t) - n_{\uparrow\downarrow}(t)$ and
\textbf{c}: doublons $n_{\uparrow\downarrow}(t)$. \textbf{d}: Pair correlation function
$n_k$ for the ground state. The peak in $n_k$ and $q$ as determined from 
the edge expansion velocity of unpaired particles
give the same value $q = 0.67 \frac{1}{L}$.} 
\label{fig:bigparams} 
\end{figure}

Looking at Figure \ref{fig:q_trap_all}, the $q$ determined from
the expansion velocity matches $k_{F \uparrow} - k_{F \downarrow}$ which has been
calculated from noninteracting quantum harmonic oscillator eigenstates. 
To elaborate, for example $k_{F \uparrow}$ for five particles has been 
obtained from the peak at maximum momentum in the momentum distribution 
of the 5th quantum harmonic oscillator eigenstate (i.e.\ $n=4$, since the
harmonic oscillator quantum numbers start from $n=0$). Moreover, 
the pair momentum correlator $n_k$ matches $k_{(N_\uparrow -N_\downarrow)\uparrow} - k_{0 \downarrow}$ 
calculated again using the noninteracting harmonic oscillator eigenstates
($k_{0\downarrow} = 0$ for the 1st eigenstate which is a Gaussian centered
at zero). The quantity $k_{(N_\uparrow -N_\downarrow)\uparrow} - k_{0 \downarrow}$
corresponds to pairing between a down particle at the lowest
level ($n=0$) and an up particle at the level $n=N_\uparrow -N_\downarrow$, 
whereas $k_{F \uparrow} - k_{F \downarrow}$ corresponds to 
a pairing between a down particle at $n_{F \downarrow}$ and an up particle 
at $n_{F \uparrow}$. For larger particle numbers 
the peak of $n_k$ and $k_{F \uparrow} - k_{F \downarrow}$ 
(and thus $q$ as obtained from the expansion velocity) 
converge into the same value, as verified by our numerical simulatons 
with larger particle numbers (see Figure \ref{fig:bigparams}). 
In comparison, for the box potential, the peak of $n_k$, $k_{F \uparrow} - k_{F \downarrow}$ and 
$q$ as obtained from the expansion velocity are alread the same
for small particle numbers due to the box eigenstates being momentum eigenstates
(see Figure \ref{fig:q_box_all}).
For larger particle numbers, the box potential becomes a reasonable approximation
for the harmonic potential, and thus the values of
the peak of $n_k$, $k_{F \uparrow} - k_{F \downarrow}$ and
$q$ as obtained from the expansion velocity become the same.
This explains why the local density approximation works also in
a harmonic trap for determining the FFLO momentum using 
$k_{F \uparrow} - k_{F \downarrow} = n_{\uparrow} - n_{\downarrow}$ for 
large particle numbers.

\clearpage

\clearpage

\begin{thebibliography}{29}
\expandafter\ifx\csname natexlab\endcsname\relax\def\natexlab#1{#1}\fi
\expandafter\ifx\csname bibnamefont\endcsname\relax
  \def\bibnamefont#1{#1}\fi
\expandafter\ifx\csname bibfnamefont\endcsname\relax
  \def\bibfnamefont#1{#1}\fi
\expandafter\ifx\csname citenamefont\endcsname\relax
  \def\citenamefont#1{#1}\fi
\expandafter\ifx\csname url\endcsname\relax
  \def\url#1{\texttt{#1}}\fi
\expandafter\ifx\csname urlprefix\endcsname\relax\def\urlprefix{URL }\fi
\providecommand{\bibinfo}[2]{#2}
\providecommand{\eprint}[2][]{\url{#2}}

\bibitem[{\citenamefont{Anderson et~al.}(1995)\citenamefont{Anderson, Ensher,
  Matthews, Wieman, and Cornell}}]{Anderson:1995dw}
\bibinfo{author}{\bibfnamefont{M.~H.} \bibnamefont{Anderson}},
  \bibnamefont{et~al.}, \bibinfo{journal}{Science}
  \textbf{\bibinfo{volume}{269}}, \bibinfo{pages}{198} (\bibinfo{year}{1995}).

\bibitem[{\citenamefont{Andrews et~al.}(1997)\citenamefont{Andrews, Townsend,
  Miesner, Durfee, Kurn, and Ketterle}}]{Andrews:1997uf}
\bibinfo{author}{\bibfnamefont{M.~R.} \bibnamefont{Andrews}},
  \bibnamefont{et~al.}, \bibinfo{journal}{Science}
  \textbf{\bibinfo{volume}{275}}, \bibinfo{pages}{637} (\bibinfo{year}{1997}).

\bibitem[{\citenamefont{Greiner et~al.}(2002)\citenamefont{Greiner, Mandel,
  Esslinger, H{\"a}nsch, and Bloch}}]{Greiner:2002do}
\bibinfo{author}{\bibfnamefont{M.}~\bibnamefont{Greiner}},
  \bibnamefont{et~al.}, \bibinfo{journal}{Nature}
  \textbf{\bibinfo{volume}{415}}, \bibinfo{pages}{39} (\bibinfo{year}{2002}).

\bibitem[{\citenamefont{O'Hara et~al.}(2002)\citenamefont{O'Hara, Hemmer, Gehm,
  Granade, and Thomas}}]{OHara:2002br}
\bibinfo{author}{\bibfnamefont{K.}~\bibnamefont{O'Hara}}, \bibnamefont{et~al.},
  \bibinfo{journal}{Science} \textbf{\bibinfo{volume}{298}},
  \bibinfo{pages}{2179} (\bibinfo{year}{2002}).

\bibitem[{\citenamefont{Menotti et~al.}(2002)\citenamefont{Menotti, Pedri, and
  Stringari}}]{Menotti:2002ix}
\bibinfo{author}{\bibfnamefont{C.}~\bibnamefont{Menotti}},
  \bibinfo{author}{\bibfnamefont{P.}~\bibnamefont{Pedri}}, \bibnamefont{and}
  \bibinfo{author}{\bibfnamefont{S.}~\bibnamefont{Stringari}},
  \bibinfo{journal}{Phys. Rev. Lett.} \textbf{\bibinfo{volume}{89}},
  \bibinfo{pages}{250402} (\bibinfo{year}{2002}).

\bibitem[{\citenamefont{Radovan et~al.}(2003)\citenamefont{Radovan, Fortune,
  Murphy, Hannahs, Palm, Tozer, and Hall}}]{Radovan:2003gl}
\bibinfo{author}{\bibfnamefont{H.~A.} \bibnamefont{Radovan}},
  \bibnamefont{et~al.}, \bibinfo{journal}{Nature}
  \textbf{\bibinfo{volume}{425}}, \bibinfo{pages}{51} (\bibinfo{year}{2003}).

\bibitem[{\citenamefont{Bianchi et~al.}(2003)\citenamefont{Bianchi, Movshovich,
  Capan, Pagliuso, and Sarrao}}]{Bianchi:2003jm}
\bibinfo{author}{\bibfnamefont{A.}~\bibnamefont{Bianchi}},
\bibinfo{author}{\bibfnamefont{R.}~\bibnamefont{Movshovich}},
\bibinfo{author}{\bibfnamefont{C.}~\bibnamefont{Capan}},
\bibinfo{author}{\bibfnamefont{P.~G.}~\bibnamefont{Pagliuso}},
\bibinfo{author}{\bibfnamefont{J.~L.}~\bibnamefont{Sarrao}},
\bibinfo{journal}{Phys. Rev. Lett.}
  \textbf{\bibinfo{volume}{91}},
  \bibinfo{pages}{187004} (\bibinfo{year}{2003}).

\bibitem[{\citenamefont{Casalbuoni and Nardulli}(2004)}]{Casalbuoni:2004fn}
\bibinfo{author}{\bibfnamefont{R.}~\bibnamefont{Casalbuoni}} \bibnamefont{and}
  \bibinfo{author}{\bibfnamefont{G.}~\bibnamefont{Nardulli}},
  \bibinfo{journal}{Rev. Mod. Phys.} \textbf{\bibinfo{volume}{76}},
  \bibinfo{pages}{263} (\bibinfo{year}{2004}).

\bibitem[{\citenamefont{Liao et~al.}(2010)\citenamefont{Liao, Rittner,
  Paprotta, Li, Partridge, Hulet, Baur, and Mueller}}]{Liao:2010bu}
\bibinfo{author}{\bibfnamefont{Y.-A.} \bibnamefont{Liao}},
  \bibnamefont{et~al.}, \bibinfo{journal}{Nature}
  \textbf{\bibinfo{volume}{467}}, \bibinfo{pages}{567} (\bibinfo{year}{2010}).

\bibitem[{\citenamefont{Fulde and Ferrell}(1964)}]{Fulde:1964dq}
\bibinfo{author}{\bibfnamefont{P.}~\bibnamefont{Fulde}} \bibnamefont{and}
  \bibinfo{author}{\bibfnamefont{R.~A.} \bibnamefont{Ferrell}},
  \bibinfo{journal}{Phys. Rev.} \textbf{\bibinfo{volume}{135}},
  \bibinfo{pages}{A550} (\bibinfo{year}{1964}).

\bibitem[{\citenamefont{Larkin and Ovchinnikov}(1964)}]{Larkin:1964uw}
\bibinfo{author}{\bibfnamefont{A.~I.} \bibnamefont{Larkin}} \bibnamefont{and}
  \bibinfo{author}{\bibfnamefont{Y.~N.} \bibnamefont{Ovchinnikov}},
  \bibinfo{journal}{Zh. Eksp. Teor. Fiz.} \textbf{\bibinfo{volume}{47}},
  \bibinfo{pages}{1136} (\bibinfo{year}{1964}).

\bibitem[{\citenamefont{Takada and Izuyama}(1969)}]{Takada:1969fo}
\bibinfo{author}{\bibfnamefont{S.}~\bibnamefont{Takada}} \bibnamefont{and}
  \bibinfo{author}{\bibfnamefont{T.}~\bibnamefont{Izuyama}},
  \bibinfo{journal}{Prog. Theor. Phys.} \textbf{\bibinfo{volume}{41}},
  \bibinfo{pages}{635} (\bibinfo{year}{1969}).

\bibitem[{\citenamefont{Feiguin and Heidrich-Meisner}(2007)}]{Feiguin:2007if}
\bibinfo{author}{\bibfnamefont{A.~E.} \bibnamefont{Feiguin}} \bibnamefont{and}
  \bibinfo{author}{\bibfnamefont{F.}~\bibnamefont{Heidrich-Meisner}},
  \bibinfo{journal}{Phys. Rev. B} \textbf{\bibinfo{volume}{76}},
  \bibinfo{pages}{220508} (\bibinfo{year}{2007}).

\bibitem[{\citenamefont{Batrouni et~al.}(2008)\citenamefont{Batrouni, Huntley,
  Rousseau, and Scalettar}}]{Batrouni:2008fwa}
\bibinfo{author}{\bibfnamefont{G.~G.} \bibnamefont{Batrouni}},
\bibinfo{author}{\bibfnamefont{M.~H.} \bibnamefont{Huntley}},
\bibinfo{author}{\bibfnamefont{V.~G.} \bibnamefont{Rousseau}},
\bibinfo{author}{\bibfnamefont{R.~T.} \bibnamefont{Scalettar}},
\bibinfo{journal}{Phys. Rev. Lett.}
  \textbf{\bibinfo{volume}{100}}, \bibinfo{pages}{116405}
  (\bibinfo{year}{2008}).

\bibitem[{\citenamefont{Rizzi et~al.}(2008)\citenamefont{Rizzi, Polini,
  Cazalilla, Bakhtiari, Tosi, and Fazio}}]{Rizzi:2008jk}
\bibinfo{author}{\bibfnamefont{M.}~\bibnamefont{Rizzi}}, \bibnamefont{et~al.},
  \bibinfo{journal}{Phys. Rev. B} \textbf{\bibinfo{volume}{77}},
  \bibinfo{pages}{245105} (\bibinfo{year}{2008}).

\bibitem[{\citenamefont{Yang}(2001)}]{Yang:2001iz}
\bibinfo{author}{\bibfnamefont{K.}~\bibnamefont{Yang}}, \bibinfo{journal}{Phys.
  Rev. B} \textbf{\bibinfo{volume}{63}},
 \bibinfo{pages}{140511} (\bibinfo{year}{2001}).

\bibitem[{\citenamefont{Orso}(2007)}]{Orso:2007uu}
\bibinfo{author}{\bibfnamefont{G.}~\bibnamefont{Orso}}, \bibinfo{journal}{Phys.
  Rev. Lett.} \textbf{\bibinfo{volume}{98}}, \bibinfo{pages}{070402}
  (\bibinfo{year}{2007}).

\bibitem[{\citenamefont{Parish et~al.}(2007)\citenamefont{Parish, Baur,
  Mueller, and Huse}}]{Parish:2007je}
\bibinfo{author}{\bibfnamefont{M.~M.} \bibnamefont{Parish}},
\bibinfo{author}{\bibfnamefont{S.~K.} \bibnamefont{Baur}},
\bibinfo{author}{\bibfnamefont{E.~J.} \bibnamefont{Mueller}},
\bibinfo{author}{\bibfnamefont{D.~A.} \bibnamefont{Huse}},
\bibinfo{journal}{Phys. Rev. Lett.}
  \textbf{\bibinfo{volume}{99}}, \bibinfo{pages}{250403} (\bibinfo{year}{2007}).

\bibitem[{\citenamefont{Zwierlein et~al.}(2006)\citenamefont{Zwierlein,
  Schirotzek, Schunck, and Ketterle}}]{Zwierlein:2006gb}
\bibinfo{author}{\bibfnamefont{M.~W.} \bibnamefont{Zwierlein}},
  \bibnamefont{et~al.}, \bibinfo{journal}{Science}
  \textbf{\bibinfo{volume}{311}}, \bibinfo{pages}{492} (\bibinfo{year}{2006}).

\bibitem[{\citenamefont{Partridge et~al.}(2006)\citenamefont{Partridge, Li,
  Kamar, Liao, and Hulet}}]{Partridge:2006hx}
\bibinfo{author}{\bibfnamefont{G.~B.} \bibnamefont{Partridge}},
  \bibnamefont{et~al.}, \bibinfo{journal}{Science}
  \textbf{\bibinfo{volume}{311}}, \bibinfo{pages}{503} (\bibinfo{year}{2006}).

\bibitem[{\citenamefont{Bakhtiari et~al.}(2008)\citenamefont{Bakhtiari,
  Leskinen, and T{\"o}rm{\"a}}}]{Bakhtiari:2008iqa}
\bibinfo{author}{\bibfnamefont{M.~R.} \bibnamefont{Bakhtiari}},
  \bibinfo{author}{\bibfnamefont{M.~J.} \bibnamefont{Leskinen}},
  \bibnamefont{and}
  \bibinfo{author}{\bibfnamefont{P.}~\bibnamefont{T{\"o}rm{\"a}}},
  \bibinfo{journal}{Phys. Rev. Lett.} \textbf{\bibinfo{volume}{101}},
  \bibinfo{pages}{120404} (\bibinfo{year}{2008});
\bibinfo{author}{\bibfnamefont{A.}~\bibnamefont{Korolyuk}},
  \bibinfo{author}{\bibfnamefont{F.}~\bibnamefont{Massel}}, \bibnamefont{and}
  \bibinfo{author}{\bibfnamefont{P.}~\bibnamefont{T{\"o}rm{\"a}}},
  \bibinfo{journal}{Phys. Rev. Lett.} \textbf{\bibinfo{volume}{104}},
  \bibinfo{pages}{236402} (\bibinfo{year}{2010});
\bibinfo{author}{\bibfnamefont{J.~M.} \bibnamefont{Edge}},
\bibinfo{author}{\bibfnamefont{N.~R.} \bibnamefont{Cooper}},
  \bibinfo{journal}{Phys. Rev. A} \textbf{\bibinfo{volume}{81}},
  \bibinfo{pages}{063606} (\bibinfo{year}{2010});
\bibinfo{author}{\bibfnamefont{M.}~\bibnamefont{Swanson}},
  \bibinfo{author}{\bibfnamefont{Y.~L.} \bibnamefont{Loh}}, \bibnamefont{and}
  \bibinfo{author}{\bibfnamefont{N.}~\bibnamefont{Trivedi}},
  \bibinfo{journal}{arXiv:} \textbf{\bibinfo{volume}{1106.3908v1}}
  (\bibinfo{year}{2011}).

\bibitem[{\citenamefont{Zhao and Liu}(2008)}]{Zhao:2010hk}
\bibinfo{author}{\bibfnamefont{E.}~\bibnamefont{Zhao}} \bibnamefont{and}
  \bibinfo{author}{\bibfnamefont{W.~V.} \bibnamefont{Liu}},
  \bibinfo{journal}{Phys. Rev. A} \textbf{\bibinfo{volume}{78}},
  \bibinfo{pages}{063605} (\bibinfo{year}{2008}).

\bibitem[{\citenamefont{Essler et~al.}(2005)\citenamefont{Essler, Frahm,
  G{\"o}hmann, Kl{\"u}mper, and Korepin}}]{Essler:2005uw}
\bibinfo{author}{\bibfnamefont{E.H.}~\bibnamefont{Lieb}}, \bibnamefont{and}
  \bibinfo{author}{\bibfnamefont{F.Y.}~\bibnamefont{Wu}},
  \bibinfo{journal}{Phys. Rev. Lett.} \textbf{\bibinfo{volume}{20}},
  \bibinfo{pages}{1445} (\bibinfo{year}{1968});
\bibinfo{author}{\bibfnamefont{F.~H.~L.} \bibnamefont{Essler}},
  \bibnamefont{et~al.}, \emph{\bibinfo{title}{{The One-Dimensional Hubbard
  Model}}} (\bibinfo{publisher}{Cambridge University Press},
  \bibinfo{year}{2005}).

\bibitem[{\citenamefont{Daley et~al.}(2004)\citenamefont{Daley, Kollath,
  Schollwock, and Vidal}}]{Daley:2004hk}
\bibinfo{author}{\bibfnamefont{A.~J.} \bibnamefont{Daley}},
  \bibnamefont{et~al.}, \bibinfo{journal}{J Stat Mech-Theory E} p.
  \bibinfo{pages}{P04005} (\bibinfo{year}{2004}).

\bibitem[{\citenamefont{Kajala et~al.}()\citenamefont{Kajala, Massel, and
  T{\"o}rm{\"a}}}]{Kajala:8BtvNK09}
\bibinfo{author}{\bibfnamefont{J.}~\bibnamefont{Kajala}},
  \bibinfo{author}{\bibfnamefont{F.}~\bibnamefont{Massel}}, \bibnamefont{and}
  \bibinfo{author}{\bibfnamefont{P.}~\bibnamefont{T{\"o}rm{\"a}}},
  \emph{\bibinfo{title}{{Supplementary information}}}.

\bibitem[{\citenamefont{Tezuka and Ueda}(2008)}]{Tezuka:2008fp}
\bibinfo{author}{\bibfnamefont{M.}~\bibnamefont{Tezuka}} \bibnamefont{and}
  \bibinfo{author}{\bibfnamefont{M.}~\bibnamefont{Ueda}},
  \bibinfo{journal}{Phys. Rev. Lett.} \textbf{\bibinfo{volume}{100}},
  \bibinfo{pages}{110403} (\bibinfo{year}{2008}).

\bibitem[{\citenamefont{Giamarchi}(2004)}]{Giamarchi:2004vu}
\bibinfo{author}{\bibfnamefont{T.}~\bibnamefont{Giamarchi}},
  \emph{\bibinfo{title}{{Quantum Physics in One Dimension (International Series
  of Monographs on Physics)}}} (\bibinfo{publisher}{Oxford University Press,
  USA}, \bibinfo{year}{2004}).

\bibitem[{\citenamefont{Oelkers et~al.}(2006)\citenamefont{Oelkers, Batchelor,
  Bortz, and Guan}}]{Oelkers:2006hi}
\bibinfo{author}{\bibfnamefont{N.}~\bibnamefont{Oelkers}},
  \bibnamefont{et~al.}, \bibinfo{journal}{J Phys A-Math Gen}
  \textbf{\bibinfo{volume}{39}}, \bibinfo{pages}{1073} (\bibinfo{year}{2006}).

\bibitem[{\citenamefont{Wolak et~al.}(2010)\citenamefont{Wolak, Rousseau,
  Miniatura, Gremaud, Scalettar, and Batrouni}}]{Wolak:2010ib}
\bibinfo{author}{\bibfnamefont{M.}~\bibnamefont{Wolak}}, \bibnamefont{et~al.},
  \bibinfo{journal}{Phys. Rev. A} \textbf{\bibinfo{volume}{82}},
  \bibinfo{pages}{013614} (\bibinfo{year}{2010}).


\end{thebibliography}

\clearpage

\begin{thebibliography}{2}
\expandafter\ifx\csname natexlab\endcsname\relax\def\natexlab#1{#1}\fi
\expandafter\ifx\csname bibnamefont\endcsname\relax
  \def\bibnamefont#1{#1}\fi
\expandafter\ifx\csname bibfnamefont\endcsname\relax
  \def\bibfnamefont#1{#1}\fi
\expandafter\ifx\csname citenamefont\endcsname\relax
  \def\citenamefont#1{#1}\fi
\expandafter\ifx\csname url\endcsname\relax
  \def\url#1{\texttt{#1}}\fi
\expandafter\ifx\csname urlprefix\endcsname\relax\def\urlprefix{URL }\fi
\providecommand{\bibinfo}[2]{#2}
\providecommand{\eprint}[2][]{\url{#2}}

\bibitem[{\citenamefont{Oelkers et~al.}(2006)\citenamefont{Oelkers, Batchelor,
  Bortz, and Guan}}]{Oelkers:2006hi:s}
\bibinfo{author}{\bibfnamefont{N.}~\bibnamefont{Oelkers}},
  \bibnamefont{et~al.}, \bibinfo{journal}{J Phys A-Math Gen}
  \textbf{\bibinfo{volume}{39}}, \bibinfo{pages}{1073} (\bibinfo{year}{2006}).

\bibitem[{\citenamefont{Kajala et~al.}(2011)\citenamefont{Kajala, Massel, and
  T{\"o}rm{\"a}}}]{Kajala:2011ho}
\bibinfo{author}{\bibfnamefont{J.}~\bibnamefont{Kajala}},
  \bibinfo{author}{\bibfnamefont{F.}~\bibnamefont{Massel}}, \bibnamefont{and}
  \bibinfo{author}{\bibfnamefont{P.}~\bibnamefont{T{\"o}rm{\"a}}},
  \bibinfo{journal}{Phys. Rev. Lett.} \textbf{\bibinfo{volume}{106}},
  \bibinfo{pages}{206401} (\bibinfo{year}{2011}).

\end{thebibliography}
\end{document}